\documentclass[%
 reprint,
 amsmath,amssymb,
 aps,
]{revtex4-2}

\pdfoutput=1
\usepackage{graphicx}
\usepackage{tabularx}
\usepackage{rotating}
\usepackage{dcolumn}
\usepackage{bm}

\usepackage{amssymb}
\usepackage{pifont}
\usepackage{multirow}
\usepackage{mathrsfs}
\usepackage{mathtools}
\usepackage{makecell}
\usepackage{bbm}
\usepackage{amsmath}
\usepackage{bm}
\DeclarePairedDelimiter\bra{\langle}{\rvert}
\DeclarePairedDelimiter\ket{\lvert}{\rangle}
\DeclarePairedDelimiterX\braket[2]{\langle}{\rangle}{#1 \delimsize\vert #2}
\usepackage{mathtools}
\usepackage[toc,page]{appendix}
\DeclarePairedDelimiter{\abs}{\lvert}{\rvert}
\usepackage{cellspace} %
\newcommand{\Gj}[6]{ \begin{Bmatrix}
   #1 & #2 & #3 \\
   #4 & #5 & #6 
  \end{Bmatrix}}

\usepackage[utf8]{inputenc}

\usepackage[margin=1in]{geometry}
\usepackage{datetime}
\usepackage{xcolor}

\newdateformat{monthdayyeardate}{%
  \monthname[\THEMONTH]~\THEDAY, \THEYEAR}

\begin{document}

\title{Photon scattering errors during stimulated Raman transitions \linebreak in trapped-ion qubits}
\author{{I. D. Moore${}^*$}, {W. C. Campbell${}^{\dagger,\ddagger,\S}$}, {E. R. Hudson${}^{\dagger,\ddagger,\S}$}, \\ {M. J. Boguslawski}${}^{\dagger,\S}$, {D. J. Wineland${}^*$}, and {D. T. C. Allcock${}^*$}}
\affiliation{${}^*$Department of Physics, University of Oregon, Eugene, OR, USA}
\affiliation{${}^\dagger$Department of Physics and Astronomy, University of California Los Angeles, Los Angeles, CA, USA}
\affiliation{${}^\ddagger$Center for Quantum Science and Engineering, University of California Los Angeles, Los Angeles, CA, USA}
\affiliation{${}^\S$Challenge Institute for Quantum Computation, University of California Los Angeles, Los Angeles, CA, USA}

\date{\monthdayyeardate\today}
\begin{abstract}

We study photon scattering errors in stimulated Raman driven quantum logic gates.
For certain parameter regimes, we find that previous, simplified models of the process significantly overestimate the gate error rate due to photon scattering.
This overestimate is shown to be due to previous models neglecting the detuning dependence of the scattered photon frequency and Lamb-Dicke parameter, a second scattering process, interference effects on scattering rates to metastable manifolds, and the counter-rotating contribution to the Raman transition rate. 
The resulting improved model shows that there is no fundamental limit on gate error due to photon scattering for electronic ground state qubits in commonly-used trapped-ion species when the Raman laser beams are red detuned from the main optical transition. 
Additionally, photon scattering errors are studied for qubits encoded in metastable $D_{5/2}$ manifold, showing that gate errors below $10^{-4}$ are achievable for all commonly-used trapped ions.
\end{abstract}

\maketitle

\section{Introduction}\label{sec:Introduction}
Stimulated Raman transitions are often used to implement trapped-ion quantum logic gates. In such setups, a pair of laser beams drives a two-photon stimulated transition between qubit states $\ket{0}$ and $\ket{1}$ through one or more intermediate states. 
Photon scattering during this process is unavoidable and understanding the fundamental limit it places on achievable gate fidelity is potentially crucial for the viability of trapped-ion quantum computing. 

Although photon scattering errors have been previously studied~\cite{Ozeri2007, Sawyer2020}, we revisit the topic for two reasons. 
First, past work~\cite{Ozeri2007} produced a model of scattering that,  while suitable for explaining moderate-detuning gate errors, is inaccurate at larger detunings. 
This inaccuracy is due to neglecting the detuning dependence of the scattered photon frequency and Lamb-Dicke parameter, contributions of a second scattering term (the `V' scattering process in Fig.~\ref{fig:TotalScatteringDiagram}), interference effects in scattering to the metastable manifolds, and the contribution of the counter-rotating component of the laser electric field to the Raman transition rate. 
As shown below, including such effects changes the limiting behavior of the model, resulting in no lower bound on gate error, in contrast to the predictions of~\cite{Ozeri2007}.

Second, the previous studies focused solely on qubits with either one or both states encoded in the $S_{1/2}$ ground manifold of trapped ions (`$o$' qubits and `$g$' qubits, respectively).
However, a recent proposal~\cite{MetastableGeniuses} has shown that encoding qubits in metastable states of trapped ions (`$m$' qubits) has several important advantages over $g$ and $o$ type encodings. 
Therefore, we also study photon scattering errors for $m$ qubits. 

Below, after describing our qubit choices and defining our model, we calculate the photon-scattering-induced errors in single-qubit and two-qubit gates in $m$ qubits and compare them to $g$ qubits. There are many physical differences between $m$ and $g$ qubits, such as the dipole coupling between the qubit states and $P_{1/2}$ that exists for $g$ qubits but is absent in $m$ qubits, or certain scattering terms present in $m$ qubits that are not present in $g$ qubits. These physical differences exert influences of varying magnitudes on the quantitative scattering error, but on net, they tend to increase the detuning from resonance required for a certain error in $m$ qubits relative to $g$ qubits. As for the qualitative scattering behavior, we find that the main difference between the two schemes is the existence of a lower bound on two-qubit gate errors in $m$ qubits which is absent in $g$ qubits. However, for all trapped ions considered in this work, this lower bound is sufficiently small (less than 10${}^{-4}$) that low-overhead error correction is possible for Raman gates in the $m$ qubit scheme.

After describing the overall behavior of the two different qubit schemes in the large-detuning model, we discuss the contributions of higher-lying levels to the scattering rates of the $m$ qubit model. Additionally, for both $m$ and $g$ qubits, we estimate the contribution to gate error from Rayleigh scattering.

\begin{figure*}[!htb]
    \centering
    \includegraphics[width=0.90\textwidth]{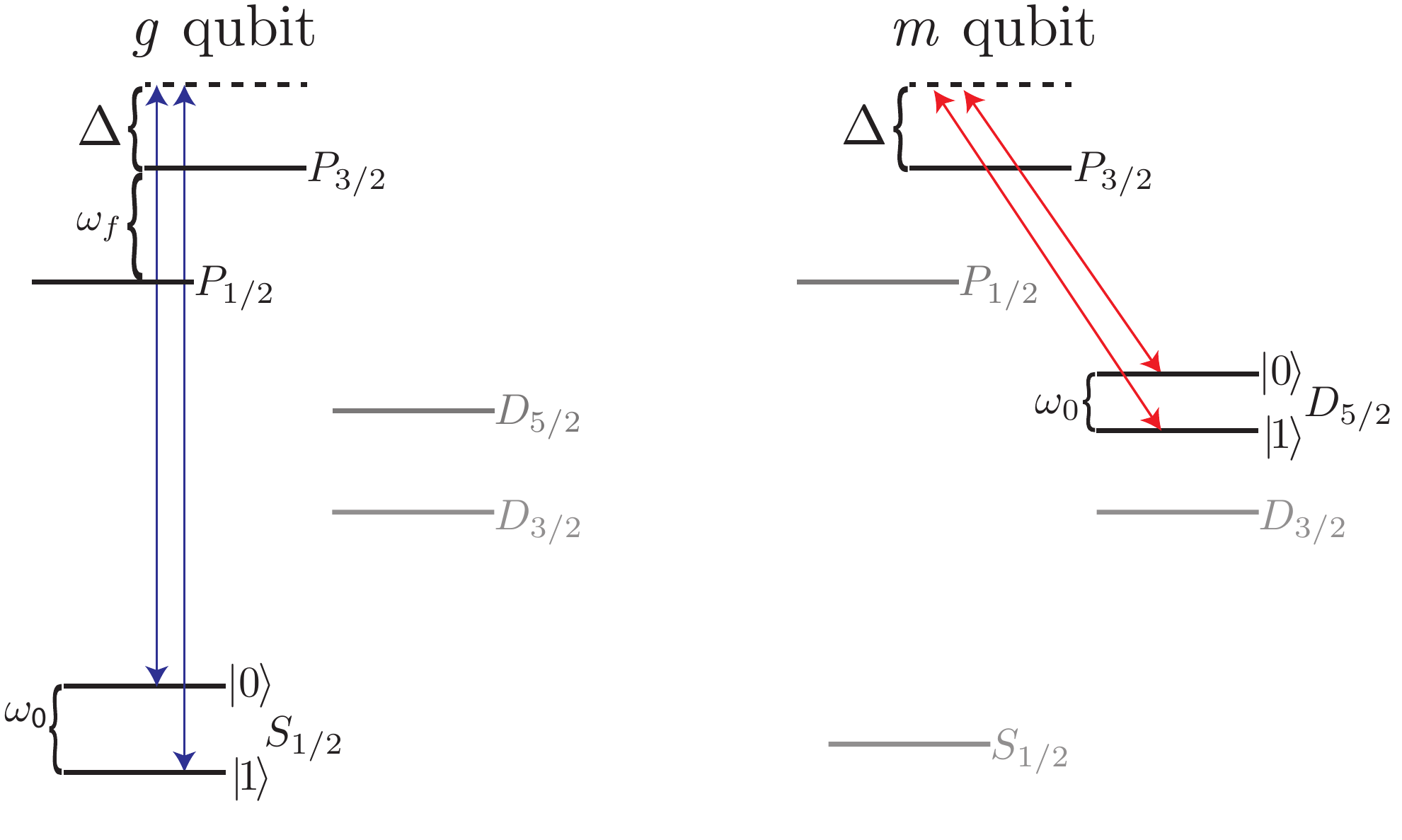}
    \caption{The \textit{g} qubits are encoded in hyperfine sublevels of the $S_{1/2}$ manifold with gates often performed via Raman transitions; \textit{m} qubits are encoded in the hyperfine sublevels of the $D_{5/2}$ level with gates performed in the same way. Manifolds that do not participate in qubit operations in the respective schemes are shown in gray.}
    \label{fig:FidelityFigs}
\end{figure*}

\section{Choice of Qubit}\label{sec:ChoiceOfQubit}
In what follows, we consider a number of common trapped-ion qubit species: ${}^{9}$Be${}^+$, ${}^{25}$Mg${}^+$, ${}^{43}$Ca${}^+$, ${}^{87}$Sr${}^+$, ${}^{133}$Ba${}^+$, ${}^{135}$Ba${}^+$, ${}^{137}$Ba${}^+$, ${}^{171}$Yb${}^+$, and ${}^{173}$Yb${}^+$. We present results for \textit{g} qubits in all species. For \textit{m} qubits, we perform calculations only for the species with a sufficiently long-lived ($\gtrsim1$ second lifetime) $D_{5/2}$ manifold: ${}^{43}$Ca${}^+$, ${}^{87}$Sr${}^+$, ${}^{133}$Ba${}^+$, ${}^{135}$Ba${}^+$, and ${}^{137}$Ba${}^+$ (see \footnote{Yb${}^+$ has an $F_{7/2}$ manifold with a years-long lifetime~\cite{MetastableGeniuses}, making it a suitable candidate for \textit{m} qubits. However, this system is more complex and would require a separate analysis, so we only present \textit{g} qubit results for Yb${}^+$.} for further discussion). The attributes of all these ion species are given in Table~\ref{table:QubitCharacterization}.

For our qubit choices, we use hyperfine `clock' qubits for both $m$ and $g$ qubits, due to their insensitivity to magnetic field noise and corresponding suppression of Rayleigh dephasing~\cite{TingRei2016} (see \footnote{A clock qubit is defined as having a qubit frequency that does not change with changes in the $B$-field (to first order); the $B$-field magnitude at which this occurs can be called the clock point, and the $B$-field direction defines the quantization axis.} for the definition of `clock qubit'.). We give our qubit choices (along with other qubit details) explicitly in Table~\ref{table:QubitCharacterization} in terms of $F$ (hyperfine angular momentum quantum number) and $m_F$ (corresponding angular momentum projection quantum number).  Since $g$ qubits are already well-established, we simply follow~\cite{Ozeri2007} in their encoding scheme. As for $m$ qubits, we choose states that should be relatively easy to prepare and readout, as well as give consistent behavior across all $m$ qubit species for studying the scattering errors the states ${\ket{0} = {\ket{D_{5/2}, F = I + \frac{5}{2}, m_F = \pm \abs{I + \frac{5}{2}-1}}}}$ and ${\ket{1} = {\ket{D_{5/2}, F = I + \frac{5}{2}-1, m_F = \pm (I + \frac{5}{2}-1)}}}$ (depending on the sign of relevant hyperfine splittings), where $F$ is the hyperfine angular momentum quantum number, $m_F$ is the corresponding angular momentum projection quantum number, and $I$ is the nuclear spin  (for further discussion of this choice, see \footnote{These are not necessarily ``optimal'' qubits. We choose them simply to have consistent behavior across $m$ qubits for studying the scattering errors.}). Details about the qubits in the various species are given in Table~\ref{table:QubitCharacterization}.

\begin{table*}[t]
\setcellgapes{1pt}\makegapedcells
\begin{tabular}{ | c | c | c | c | c | c | c | c | c | c |}

\hline
& ${}^{9}$Be${}^+$ & ${}^{25}$Mg${}^+$ & ${}^{43}$Ca${}^+$ & ${}^{87}$Sr${}^+$ & ${}^{133}$Ba${}^+$ & ${}^{135}$Ba${}^+$ & ${}^{137}$Ba${}^+$ & ${}^{171}$Yb${}^+$ & ${}^{173}$Yb${}^+$\\
\hline
$I$ & 3/2 & 5/2 & 7/2 & 9/2 & 1/2 & 3/2 & 3/2 & 1/2 & 5/2 \\ 
\hline
\textcolor{white}{$\tau^{D^{5/2}}$} $\tau_{D_{5/2}}$(s) \textcolor{white}{$\tau^{D^{5/2}}$} & - & - & 1.110 & 0.357 & \multicolumn{3}{c}{29.856} \vline & \multicolumn{2}{c}{0.0072}  \vline\\
\hline
$\ket{0}$ (\textit{m}) & - & - & $\ket{6, +5}$ & $\ket{6, -6}$ & $\ket{2, +2}$ & $\ket{4, -3}$ & $\ket{4, -3}$ & - & - \\ 
\hline
$\ket{1}$ (\textit{m}) & - & - & $\ket{5, +5}$ & $\ket{7, -6}$ & $\ket{3, +2}$ & $\ket{3, -3}$ & $\ket{3, -3}$ & - & - \\ 
\hline
$\ket{0}$ (\textit{g}) & $\ket{2, 0}$ & $\ket{3, 0}$ & $\ket{4, 0}$ & $\ket{5, 0}$ & $\ket{1, 0}$ & $\ket{2, 0}$ & $\ket{2, 0}$ & $\ket{0, 0}$ & $\ket{2, 0}$ \\ 
\hline
$\ket{1}$ (\textit{g}) & $\ket{1, 0}$ & $\ket{2, 0}$ & $\ket{3, 0}$ & $\ket{4, 0}$ & $\ket{0, 0}$ & $\ket{1, 0}$ & $\ket{1, 0}$ & $\ket{1, 0}$ & $\ket{3, 0}$ \\ 
\hline
Clock point (\textit{m}, G) & - & - & 2.54  & 6.49  & 33.0  & 1.79  & 0.0720  & - & - \\ 
\hline
$\omega_0/2 \pi$ (\textit{m}, GHz) & - & - & 0.025 & 0.036 & 0.062 & 0.012 & 0.00047 & - & - \\
\hline
$\omega_0/2 \pi$ (\textit{g}, GHz) & 1.3\ & 1.8\ & 3.2\ & 5.0\ & 9.9\ & 7.2\ & 8.0\ & 12.6\ & 10.5\ \\
\hline
$d^2(\omega_0 / 2\pi)/dB^2$ ($m$, kHz/G${}^2$) & - & - & 55.9 & 36.7 & 10.6 & 119 & 1.72& - & - \\
\hline
$d^2(\omega_0 / 2\pi)/dB^2$ ($g$, kHz/G${}^2$) & 3.13 & 2.19 & 1.21 & 0.783 & 0.395 & 0.545 & 0.487& 0.309 & 0.373 \\
\hline
$\gamma_{P_{3/2}} / 2 \pi$ (MHz) & 19.4 & 41.8 & 23.2 & 24.0 & \multicolumn{3}{c}{25.2} \vline & \multicolumn{2}{c}{25.9} \vline \\
\hline
\textcolor{white}{$\alpha^{D^{5/2}}$} $\alpha_{S_{1/2}}$ \textcolor{white}{$\alpha^{D^{5/2}}$} & 1 & 1 & 0.9350 & 0.9406 & \multicolumn{3}{c}{0.7417} \vline& \multicolumn{2}{c}{0.9875} \vline \\
\hline
\textcolor{white}{$\alpha^{D^{5/2}}$} $\alpha_{D_{3/2}}$ \textcolor{white}{$\alpha^{D^{5/2}}$} & - & - & 0.0063 & 0.0063 & \multicolumn{3}{c}{0.02803} \vline& \multicolumn{2}{c}{0.0017} \vline \\
\hline
\textcolor{white}{$\alpha^{D^{5/2}}$} $\alpha_{D_{5/2}}$ \textcolor{white}{$\alpha^{D^{5/2}}$} & - & - & 0.0587 & 0.0531 & \multicolumn{3}{c}{0.2303} \vline& \multicolumn{2}{c}{0.0108} \vline \\
\hline
$\omega_f/2 \pi$ (THz) & 0.198 & 2.75 & 6.68 & 24.0 & \multicolumn{3}{c}{57.2} \vline & \multicolumn{2}{c}{99.8} \vline \\
\hline
\end{tabular}
\caption{Characteristics of the qubits and ion species we consider. Throughout the table, $m$ and $g$ denote values for $m$ or $g$ qubits. $I$ is the nuclear spin; $\tau_{D_{5/2}}$ is the $D_{5/2}$ lifetime; $\ket{0}$ and $\ket{1}$ are the qubit states in the notation $\ket{F, m_F}$; $\omega_0$ is the qubit frequency; $d^2 \omega_0/dB^2$ is the second-order $B$-field-dependence of the qubit frequency; $\gamma_{P_{3/2}}$ is the decay rate of the $P_{3/2}$ manifold; $\alpha_M$ is the branching ratio of $P_{3/2}$ to manifold $M$; and $\omega_f/2 \pi$ is the fine-structure separation of the $P$ manifolds. Lifetimes taken from \cite{Sahoo2006, Taylor1997}; $\gamma_{P_{3/2}}$ values were taken/calculated from \cite{Poulsen1975, Ansbacher1989, Gosselin1988, Safranova2010, Pinnington1995, Pinnington1997}; $\alpha_M$ values taken from \cite{Song2019, Zhang2016Sr, Zhang2020, Feldker2018}; $\omega_f$ values taken from \cite{Ozeri2007}; other values calculated from atomic parameters.}\end{table*}\label{table:QubitCharacterization}

\section{Scattering Probability}\label{sec:TotalScattering}
We assume that gates are performed on \textit{m} and \textit{g} qubits using stimulated Raman transitions (Fig.~\ref{fig:FidelityFigs}). These are coherent two-photon processes where population is transferred between $\ket{0}$ and $\ket{1}$ virtually through higher energy intermediate states $\ket{k}$. During this process, there is a chance of spontaneous photon scattering, changing the ion's state or qubit phase and causing an error. The scattering probability is therefore important for characterizing the fidelity achievable by logic gates, as it sets an upper bound.

\begin{figure*}[!htb]
    \centering
    \includegraphics[width=0.8\textwidth]{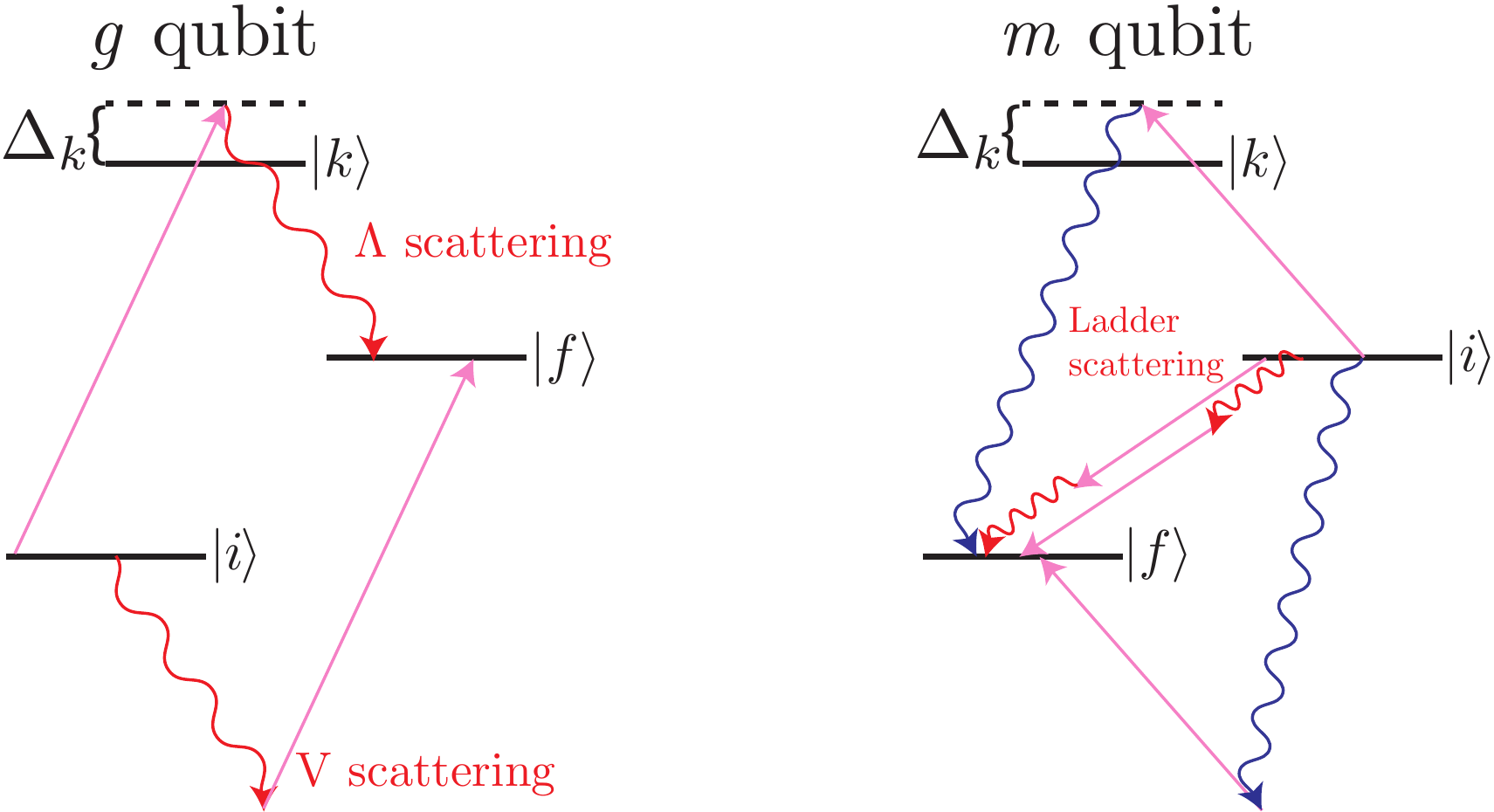}
    \caption{All two-photon scattering pathways from state $\ket{i}$ to $\ket{f}$ via excited state $\ket{k}$ in \textit{g} qubits and \textit{m} qubits. Raman laser beam shown in pink, scattered light shown in red and blue. Processes in which the Raman beam points up in the diagram are $\Lambda$ or \text{V} scattering events. Processes in which the Raman beam points down are ladder scattering events (note that ladder scattering events are allowed only in \textit{m} qubits).}
    \label{fig:TotalScatteringDiagram}
\end{figure*}

The rate of `$\Lambda$ and $\text{V}$ scattering' (the processes with upward pointing laser beam arrows in Fig.~\ref{fig:TotalScatteringDiagram}) from qubit states $\ket{i}$ to final state $\ket{f}$ can be calculated from (see Appendix~\ref{appendix:Derivation} for derivation)

\begin{widetext}
\begin{equation} \label{eqn:TotalScatteringTwoProcs_Spon}
\Gamma_{f,\Lambda\!\text{V}} = \sum_{i, j, q} \frac{e^2 E_j^2 \mu_{Pi}^2}{4 \hbar^2} \xi_i \gamma_{Pf} \abs*{\sum_{k} \left(\frac{\bra{f} \vec{r} \boldsymbol{\cdot} \mathbf{\hat{e}}_q \ket{k}\bra{k} \vec{r} \boldsymbol{\cdot} \hat{\epsilon}_j \ket{i}}{\mu_{Pf} \mu_{Pi} \left(\omega_{kP} - \Delta \right)} + \frac{\bra{f} \vec{r} \boldsymbol{\cdot} \hat{\epsilon}_j \ket{k}\bra{k} \vec{r} \boldsymbol{\cdot} \mathbf{\hat{e}}_q \ket{i}}{\mu_{Pf} \mu_{Pi} \left(\omega_{ki} + \omega_{kf} + \Delta \right)}\right)}^2 \left(1 + \frac{\Delta}{\omega_{Pf}} \right)^3,
\end{equation}
\end{widetext}
where $\xi_i$ is the occupation probability of qubit state $\ket{i}$ (on average $\frac{1}{2}$ for each qubit state in the gates we consider), $\vec{r}$ is the position operator of the electron relative to the atomic core (note that the matrix elements of the electric dipole operator can be tricky to calculate; see \cite{King2008} for a simple introduction to computing them while avoiding pitfalls), $\gamma_{Pf}$ is the decay rate from $P_{3/2}$ to the manifold containing $f$, the various $\omega_{nm}$ are transition frequencies with $n$ and $m$ corresponding to states or levels (see \footnote{To be explicit: $\omega_{kP}$ is $(E_k - E_P)/\hbar$, where $E_k$ is the mean energy of the manifold containing $\ket{k}$ and $E_P$ is the mean energy of the $P_{3/2}$ manifold (note this is zero if $k$ corresponds to $P_{3/2}$); $\omega_{ki}$ corresponds to transitions between the manifold containing $k$ and the qubit manifold; $\omega_{kf}$ corresponds to transitions between the manifold containing $k$ and the manifold containing $f$; $\omega_{Pf}$ corresponds to transitions between the $P_{3/2}$ manifold and the manifold containing $f$.} for more thorough definition of $\omega_{nm}$), $\Delta$ is the detuning measured from the $P_{3/2}$ manifold (positive for detuning above this manifold, and neglecting the energy spread of the manifold's sublevels), $E_j$ and $\hat{\epsilon}_j$ are the electric field and polarization direction of beam $j$ respectively, and $\mathbf{\hat{e}}_q$ is the scattered photon polarization. The parameters $\mu_{Pi}$ (corresponding to transitions between the $P_{3/2}$ manifold and the manifold containing state $i$) and $\mu_{Pf}$ (corresponding to transitions between the manifolds containing $k$ and $f$) are the transition dipole matrix elements of the spin-orbital coupling. Their general form can be derived by invoking the Wigner-Eckart theorem~\cite{Brink1968}, giving

\begin{widetext}
\begin{equation} \label{eqn:mudef}
\mu_{ul} = \abs*{\bra{L_l} |\vec{r}| \ket{L_u} \sqrt{(2 J_l + 1)(2 L_l + 1)} \Gj{L_l}{L_u}{1}{J_u}{J_l}{S}},
\end{equation}
\end{widetext}
where $\bra{L_u} |\vec{r}| \ket{L_l}$ is the reduced dipole matrix element for transitions between upper level $u$ and lower level $l$, $S$ is the spin, $L$ is the orbital angular momentum, $J$ is the total angular momentum ($L + S$), and the bracketed term is the Wigner 6j-symbol. The matrix element $\mu_{ul}$ can also be related to the decay rate  $\gamma_{ul}$ from the $J_u, L_u$ level to the $J_l, L_l$ level by~\cite{Steck2001}

\begin{equation} \label{eqn:lifetime}
\gamma_{ul} = \frac{e^2 \omega_{ul}^3}{3 \pi \epsilon_0 \hbar c^3} \mu_{ul}^2,
\end{equation}
where $\omega_{ul}$ is the transition frequency, and $e$ is the charge of the electron. In \textit{m} qubits (but not \textit{g} qubits), `ladder scattering' (the processes with downward pointing laser beam arrows in Fig.~\ref{fig:TotalScatteringDiagram}) can also contribute to the error. The ladder scattering rate is given by
\begin{widetext}
\begin{equation} \label{eqn:TotalScatteringTwoProcs_Stim}
\begin{split}
\Gamma_{f,\mathrm{lad}} = \frac{e^2 E_j^2 \mu_{Pi}^2}{4 \hbar^2} \sum_{i, j, q} \xi_i \gamma_{Pf} \bigg| \sum_{k} & \biggl(\frac{\bra{f} \vec{r} \boldsymbol{\cdot} \mathbf{\hat{e}}_q \ket{k}\bra{k} \vec{r} \boldsymbol{\cdot} \hat{\epsilon}_j^* \ket{i}}{\mu_{Pf} \mu_{Pi} \left(\omega_{kP} + \Delta + 2 \omega_{PD} \right)}  \\ & +  \frac{\bra{f} \vec{r} \boldsymbol{\cdot} \hat{\epsilon}_j^* \ket{k}\bra{k} \vec{r} \boldsymbol{\cdot} \mathbf{\hat{e}}_q \ket{i}}{\mu_{Pf} \mu_{Pi} \left(\omega_{kP} - \Delta + \omega_{Df} \right)}\biggr)\bigg|^2 \left(1 - \frac{2 \omega_{PD} + \Delta}{\omega_{Pf}} \right)^3.
\end{split}
\end{equation}
\end{widetext}
Note that in Eqn.\,\ref{eqn:TotalScatteringTwoProcs_Spon}, the $\Lambda$ scattering term (first term in the sum within the modulus) dominates over V scattering; similarly, in Eqn.\,\ref{eqn:TotalScatteringTwoProcs_Stim}, the first term in the modulus dominates. In both equations, these terms correspond to a two-photon scattering process where the laser photon is first absorbed or emitted, while the other weaker terms (what we call the counter-rotating contributions to the scattering rate) correspond to processes where the scattered photon is emitted first (Fig.\,\ref{fig:TotalScatteringDiagram}).

We can rewrite Eqns.~\ref{eqn:TotalScatteringTwoProcs_Spon} and \ref{eqn:TotalScatteringTwoProcs_Stim} in terms of $g_{ul}$ given in~\cite{Ozeri2007} as

\begin{equation} \label{eqn:gdef}
g_{ul} = \frac{e E \mu_{ul}}{2 \hbar}.
\end{equation}
 Note that in the following calculations, for \textit{m} qubits, $g_{Pi}$ and $\mu_{Pi}$ will correspond to transitions between the $D_{5/2}$ and $P_{3/2}$ manifolds, \textit{i.e.}, $u$ and $l$ in Eqns.~\ref{eqn:mudef} and \ref{eqn:gdef} will correspond to $P_{3/2}$ and $D_{5/2}$, respectively. For \textit{g} qubits, $g_{Pi}$ and $\mu_{Pi}$ will correspond to transitions between $S_{1/2}$ and $P_{3/2}$. Assuming that $g_{Pi}$ is the same for both Raman beams, we can now write Eqns.~\ref{eqn:TotalScatteringTwoProcs_Spon} and \ref{eqn:TotalScatteringTwoProcs_Stim} as
\begin{widetext}
\begin{equation} \label{eqn:TotalScatteringTwoProcs2_Spon}
\Gamma_{f,\Lambda\!\text{V}} = g_{Pi}^2 \sum_{i, j, q} \xi_i \gamma_{Pf} \abs*{\sum_{k} \left(\frac{\bra{f} \vec{r} \boldsymbol{\cdot} \mathbf{\hat{e}}_q \ket{k} \bra{k} \vec{r} \boldsymbol{\cdot} \hat{\epsilon}_j \ket{i}}{\mu_{Pf} \mu_{Pi}\left(\omega_{kP} - \Delta \right)} + \frac{\bra{f} \vec{r} \boldsymbol{\cdot} \hat{\epsilon}_j \ket{k} \bra{k} \vec{r} \boldsymbol{\cdot} \mathbf{\hat{e}}_q \ket{i}}{\mu_{Pf} \mu_{Pi}\left(\omega_{ki} + \omega_{kf} + \Delta \right)}\right)}^2 \left(1 + \frac{\Delta}{\omega_{Pf}} \right)^3
\end{equation}
and
\begin{equation}
\label{eqn:TotalScatteringTwoProcs2_Stim}
\begin{split}
\Gamma_{f,\mathrm{lad}} = g_{Pi}^2 \sum_{i, j, q} \xi_i \gamma_{Pf} \bigg|\sum_{k} & \biggl(\frac{\bra{f} \vec{r} \boldsymbol{\cdot} \mathbf{\hat{e}}_q \ket{k} \bra{k} \vec{r} \boldsymbol{\cdot} \hat{\epsilon}_j^* \ket{i}}{\mu_{Pf} \mu_{Pi}\left(\omega_{kP} + \Delta + 2 \omega_{PD} \right)} \\ &  + \frac{\bra{f} \vec{r} \boldsymbol{\cdot} \hat{\epsilon}_j^* \ket{k} \bra{k} \vec{r} \boldsymbol{\cdot} \mathbf{\hat{e}}_q \ket{i}}{\mu_{Pf} \mu_{Pi}\left(\omega_{kP} - \Delta + \omega_{Df} \right)}\biggr) \bigg|^2 \left(1 - \frac{2 \omega_{PD} + \Delta}{\omega_{Pf}} \right)^3,
\end{split}
\end{equation}
\end{widetext}
where $\omega_{Df}$ is the frequency of the transition between $D_{5/2}$ and the manifold containing the state $f$. Eqn.~\ref{eqn:TotalScatteringTwoProcs2_Spon} gives what we will call the ``full model" for \textit{g} qubits, and the sum of Eqns.~\ref{eqn:TotalScatteringTwoProcs2_Spon} and \ref{eqn:TotalScatteringTwoProcs2_Stim} gives the corresponding full model for \textit{m} qubits. In $m$ qubits, we include contributions to the gate error from some of the closest higher energy intermediate states, but we neglect such contributions in $g$ qubits for reasons given in section~\ref{sec:HigherLevels} below.

We will also define what we call our ``simplified model", in which we neglect the detuning dependence of the scattered photon frequency (\textit{i.e.}, we assume $(1 + \Delta/\omega_{Pf})^3 \approx 1$), neglect the contributions of the higher energy intermediate manifolds, and assume that only the first term in the squared modulus of Eqn.~\ref{eqn:TotalScatteringTwoProcs2_Spon} appreciably contributes to the scattering rate in both \textit{m} qubits and \textit{g} qubits. This model results in a simpler version of Eqn.~\ref{eqn:TotalScatteringTwoProcs2_Spon}~\cite{Wineland2003},

\begin{equation} \label{eqn:TotalScattering}
\Gamma_{f} \approx g_{Pi}^2 \sum_{i, j, q} \xi_i \gamma_{Pf} \abs*{\sum_{k} \frac{\bra{f} \vec{r} \boldsymbol{\cdot} \mathbf{\hat{e}}_q \ket{k} \bra{k} \vec{r} \boldsymbol{\cdot} \hat{\epsilon}_j \ket{i}}{\mu_{Pf} \mu_{Pi}\left(\omega_{kP} - \Delta \right)}}^2.
\end{equation}
These assumptions have also been made in studies of \textit{g} qubits (\textit{e.g.},~\cite{Ozeri2007}) where the $10^{-4}$ gate error threshold required detunings on the order of 10\,THz. However, as we will show, even at these modest detunings, our model can give large corrections to this simplified model. Despite this, the simplified model in \textit{m} qubits provides an intuitive illustration of the scattering behavior for a large range of detunings, so we therefore elect to present both models of $m$ qubits throughout this paper.

\subsection{Single-qubit gates}\label{sec:SingleQ}
We follow~\cite{Ozeri2007} in their choice of a representative single-qubit gate: a $\pi$-rotation around the $x$-axis of the equivalent Bloch sphere, a $\hat{\sigma}_x$ gate. We will assume the gates are driven by two laser beams that induce two-photon stimulated Raman transitions. In the case of \textit{m} qubits, we assume both beams are purely $\pi$-polarized because we found that they minimize the scattering probability and power requirements. In \textit{g} qubits, we assume each beam has equal parts $\sigma^+$ and $\sigma^-$ polarization. 
For this gate, the required time is given by~\cite{Ozeri2007}

\begin{equation}\label{eqn:1qGateTime}
    \tau_{1q} = \frac{\pi}{2 \abs{\Omega_R}},
\end{equation}
where the Rabi frequency $\Omega_R$ is calculated according to

\begin{figure*}[!htb]
    \centering
    \includegraphics[width=0.89\textwidth]{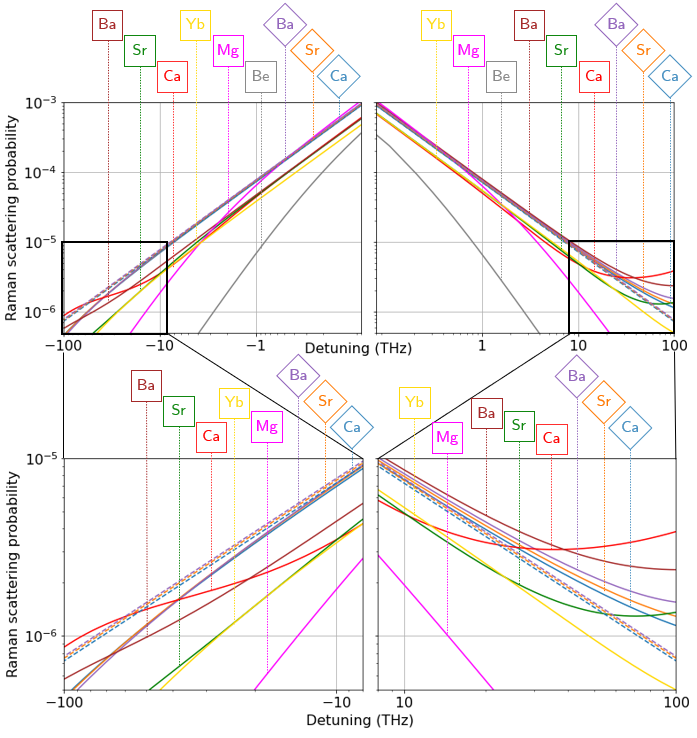}
    \caption{Raman scattering probability for \textit{m} qubits and \textit{g} qubits during single-qubit $\hat{\sigma}_x$ gate for large detuning. The \textit{m} qubit ions are labeled by diamonds and the \textit{g} qubit ions are labeled by squares. Detuning is measured relative to $P_{3/2}$ in $m$ qubits; for $g$ qubits, red detuning is measured relative to $P_{1/2}$, and blue detuning is measured relative to $P_{3/2}$. The lower plot shows zoomed-in regions of the upper plot with the simplified model behavior shown for \textit{m} qubits as dashed lines.
    }
   \label{fig:TotPlots}
\end{figure*}

\begin{widetext}
\begin{equation} \label{eqn:OR}
\Omega_R = g_{Pi}^2 \sum_k \left(\frac{\langle 1 \rvert \vec{r} \boldsymbol{\cdot} \hat{\epsilon}_r^* \lvert k \rangle \langle k \rvert \vec{r} \boldsymbol{\cdot} \hat{\epsilon}_b \lvert 0 \rangle}{\mu_{ki}^2 (\omega_{kP} - \Delta)} + \frac{\langle 1 \rvert \vec{r} \boldsymbol{\cdot} \hat\epsilon_r \lvert k \rangle \langle k \rvert \vec{r} \boldsymbol{\cdot} \hat{\epsilon}_b^* \lvert 0 \rangle}{\mu_{ki}^2 (\omega_{ki} + \omega_{Pi} + \Delta)} \right),
\end{equation}
\end{widetext}
where $r$ and $b$ (for red and blue) label the two Raman beams, $\ket{0}$ and $\ket{1}$ are the two qubit states and $k$ indexes the available intermediate states. Since Raman scattering (scattering events for which ${\ket{i} \neq \ket{f}}$) is the dominant source of errors for most species, we will, for the moment, neglect errors caused by Rayleigh scattering (scattering events for which ${\ket{i} = \ket{f}}$); Section~\ref{sec:OtherRayleigh} below gives a discussion of these errors.  We can now calculate the general Raman scattering error as

\begin{equation} \label{eqn:PRamSingleQ}
P_\mathrm{Ram} = \tau_{1q} \Gamma_\mathrm{Ram} = \frac{\pi \Gamma_\mathrm{Ram}}{2 |\Omega_R|},
\end{equation}
where $\Gamma_\mathrm{Ram}$ is given by the sum of the scattering rates to all possible final states $\ket{f}$ for which ${\ket{f} \neq \ket{i}}$. For $g$ qubits, this will be the sum of Eqn.~\ref{eqn:TotalScatteringTwoProcs2_Spon} over all relevant final states; for $m$ qubits, it will be the sum of both Eqns.~\ref{eqn:TotalScatteringTwoProcs2_Spon} and \ref{eqn:TotalScatteringTwoProcs2_Stim}  over all relevant final states. This scattering error is plotted for $g$ and $m$ qubits in Fig.~\ref{fig:TotPlots}.

In the simplified model of $m$ qubits, we neglect intermediate manifolds aside from the lowest energy $P_{3/2}$ manifold, the detuning dependence of the scattered photon frequency, ladder decay, and the counter-rotating field contribution to the scattering rate and Rabi frequency (the second terms in the sum of Eqn.~\ref{eqn:TotalScatteringTwoProcs2_Spon} and Eqn.~\ref{eqn:OR}). In this case, the sum over $k$ has the same form for all ion species considered, giving an expression for the \textit{m} qubit simplified model's Rabi frequency for each ion species the form

\begin{equation} \label{eqn:OR2}
\Omega_R = -\frac{2}{15} \frac{g_{Pi}^2}{ \Delta}.
\end{equation}
and a Raman scattering rate of the form

\begin{equation} \label{eqn:Gammacalcpi}
\Gamma_\mathrm{Ram} = \rho \frac{4}{15} \frac{g_{Pi}^2 \gamma}{\Delta^2} = 2 \rho \left| \Omega_R \frac{\gamma}{\Delta} \right| ,
\end{equation}
where $\rho$ is the Raman-only fraction of the total scattering rate (the value of $\rho$ can be inferred from the nearly constant ratio in the low-detuning regime of Fig.~\ref{fig:eray_eram}) and $\gamma$ is the decay rate of the $P_{3/2}$ manifold. For the $m$ qubit's simplified model, this results in a gate error of

\begin{equation} \label{eqn:Ptotpi}
P_\mathrm{Ram} = \rho \frac{\pi \gamma}{|\Delta|}.
\end{equation}
This simplified model of $m$ qubit scattering gives an especially simple form for $P_\mathrm{Ram}$, since only one manifold ($P_{3/2}$) appreciably contributes to the scattering. The $m$ qubit's simplified model behavior is also shown in Fig.~\ref{fig:TotPlots} as the dashed lines. 

From Fig.~\ref{fig:TotPlots}, we can see that the \textit{m} qubits and \textit{g} qubits each, as groups, have markedly similar behavior because their $P_{3/2}$ decay rates are all within $\sim$5\,MHz (Table~\ref{table:QubitCharacterization}) of each other. As expected, the full model results deviate from the initial linear regime as the detuning becomes large. In \textit{m} qubits, the full model yields a lower scattering probability than the simplified model for red-detuning, and a higher scattering probability for blue-detuning. The deviations from the \textit{m} simplified model can be observed in the lower plot of Fig.~\ref{fig:TotPlots}

\subsection{Two-qubit gates}\label{sec:TwoQ}
Consider now a two-qubit M{\o}lmer-S{\o}rensen gate, driven by three Raman beams~\cite{Molmer1999,TingRei2016}. We will suppose that the three beams are comprised of one pair of co-propagating beams of power $\mathscr{P}$ and one beam counter-propagating with this pair with intensity $2\mathscr{P}$, since this distribution of power minimizes the scattering error for this beam geometry. For this gate, the duration is

\begin{equation}\label{2qGateTime}
    \tau_{2q} = \frac{\pi}{2 \sqrt{2} \abs{\Omega_R}} \frac{\sqrt{K}}{\eta}.
\end{equation}
This is the gate time for the two-qubit gate in~\cite{Ozeri2007} but reduced by a factor of {$1/\sqrt{2}$} because of the unequal distribution of intensities in this setup. In this equation, $K$ is the number of loops the $\ket{01}$ and $\ket{10}$ states trace out in phase space (we set $K = 1$ for our calculations) and $\eta$ is the Lamb-Dicke parameter, which for counter-propagating Raman beams is given~\cite{Ozeri2007} by

\begin{equation}\label{eqn:Eta}
    \eta = \Delta k z_0 b_p^{(i)} = 2 k_L z_0 \frac{1}{\sqrt{2}} = \sqrt{2} \frac{\omega_L}{c} z_0,
\end{equation}
where $\Delta k$ is the magnitude of the difference between the two Raman beams' wavevectors (this difference wavevector being aligned to the mode of interest), $k_L$ and $\omega_L$ are respectively the wavenumber and frequency of the Raman beams, and $b_p^{(i)}$ is the mode participation factor (equal to $1/\sqrt{2}$ here). Note that the simplified model of~\cite{Ozeri2007} considers perpendicular beams which would increase the Lamb-Dicke parameter by a factor of $\sqrt{2}$. The root-mean-square spatial spread of the ground state wavefunction $z_0$ is given by

\begin{equation}
    z_0 = \sqrt{\hbar / 2 M \omega_\mathrm{trap}},
\end{equation}
where $M$ is the mass of each ion and $\omega_\mathrm{trap}$ is the frequency of the driven motional mode.

\begin{figure*}[!htb]
    \centering
    \includegraphics[width=1.0\textwidth]{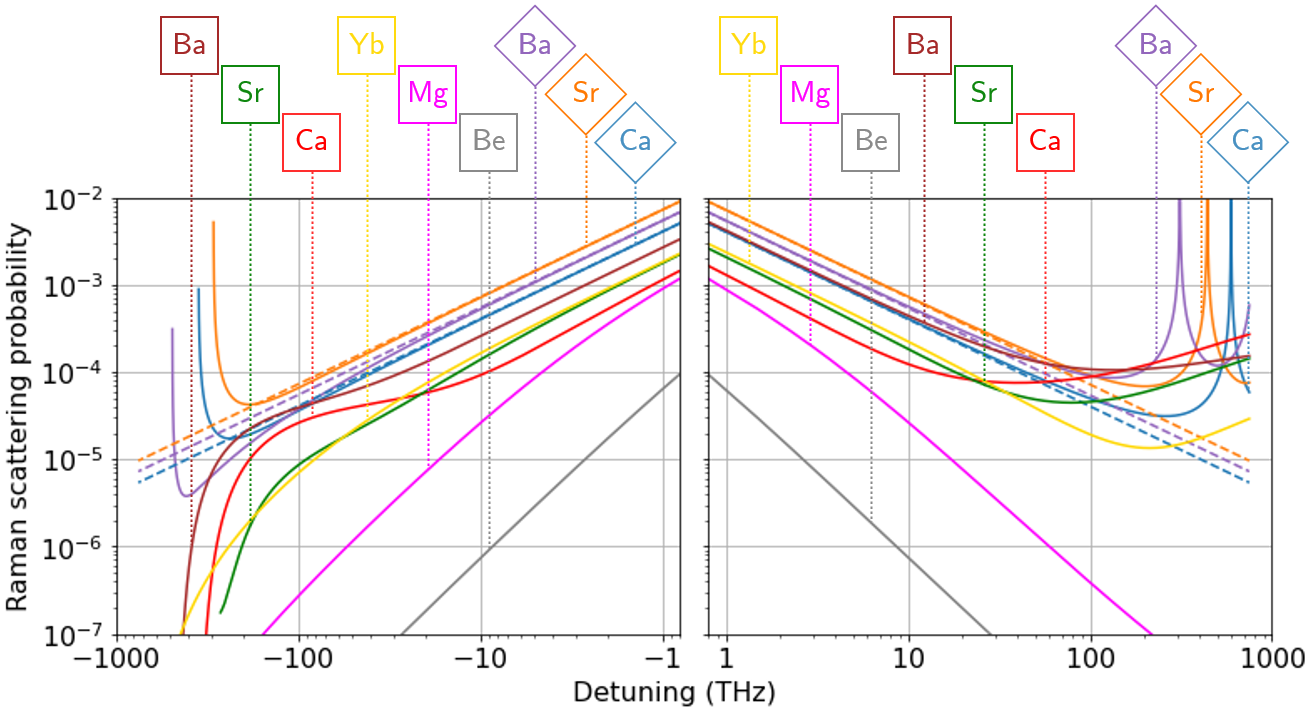}
    \caption{Raman scattering probability from a 1-loop two-qubit M{\o}lmer-S{\o}rensen gate on \textit{m} and \textit{g} qubits. The \textit{m} qubit ions are labeled by diamonds and the \textit{g} qubit ions are labeled by squares. The simple model for $m$ qubits is shown in the plot as dashed lines. Detuning is measured relative to $P_{3/2}$ in $m$ qubits; for $g$ qubits, red detuning is measured relative to $P_{1/2}$, and blue detuning is measured relative to $P_{3/2}$.}
    \label{fig:TwoQScatteringDouble}
\end{figure*}

The scattering probability for this gate is the single-qubit gate scattering probability scaled by a factor of 4, as well as the extra gate time factor~\cite{Ozeri2007}. The factor of 4 comes from two considerations: first, that the two-qubit gate uses three beams with a total power twice as great as in the total beam power used in the single-qubit gate, and second that there are two ions. Both of these differences scales the gate error by a factor of 2 to generate an overall increase of 4.  This gives a general form for the two-qubit gate Raman scattering error,

\begin{equation} \label{eqn:TwoQScattering}
    P_{R2q} = \frac{\pi \Gamma_\mathrm{Ram}}{2 \sqrt{2} |\Omega_R|} \frac{4 \sqrt{K}}{\eta}.
\end{equation}
For the simplified model in $m$ qubits, this results in

\begin{equation} \label{eqn:TwoQScattering}
    P_{R2q,simp} = \rho \frac{\pi \gamma}{|\Delta|} \frac{4 \sqrt{K}}{\sqrt{2} \eta}.
\end{equation}
Because $\eta$ is proportional to the laser frequency, it will in general depend on detuning. For the simplified model, we neglect this detuning dependence, so the error probability again exhibits linear behavior. We do, however, include this dependence in the full model, as the size of the effect is too large to reasonably ignore. In the full model for both $g$ and $m$ qubits, we can see that the Lamb-Dicke parameter leads to a further detuning dependence of the form

\begin{equation} \label{eqn:TwoQScattering2}
    P_{R2q} \propto  \frac{\omega_{Pi}}{\abs{\Delta(\omega_{Pi} + \Delta)}},
\end{equation}
where $\omega_{Pi}$ denotes the frequency of the transition between the $P_{3/2}$ manifold and the manifold containing the qubit states. The Raman scattering error of a 1-loop two-qubit M{\o}lmer-S{\o}rensen gate is plotted in Fig.~\ref{fig:TwoQScatteringDouble}; the full models of $g$ and $m$ qubits are shown by the solid curves, and the simplified $m$ qubit model is shown by the dashed curves.

The qualitative behavior of the sharply-increasing red-detuned \textit{m} qubit scattering probability in Fig.~\ref{fig:TwoQScatteringDouble} can be understood by observing that as the laser frequency approaches zero, the gate time goes to infinity (because the Lamb-Dicke parameter is approaching zero); however, the frequency of photons scattered to, \textit{e.g.}, $S_{1/2}$ approaches a non-zero value (the $S_{1/2} \leftrightarrow D_{5/2}$ transition frequency). The gate error, being the product of the scattering rate and gate time, therefore approaches infinity as the laser frequency goes to zero. This does not occur in \textit{g} qubits, because the scattered photon frequency also goes to zero as the laser frequency approaches zero. Additionally, the scattering rate in $g$ qubits is generally lower than in $m$ qubits across the detuning range considered. However, scattering errors in $m$ qubits are largely due to scattering to $S_{1/2}$, \textit{i.e.}, outside the qubit manifold. Such errors are easier to detect and correct; indeed, one of us (WCC) recently applied some of the results presented here in an investigation of erasure conversion in quantum error correction with metastable states~\cite{Kang2022}.

The form of the \textit{g} scattering probability plotted in Figs.~\ref{fig:TotPlots} and~\ref{fig:TwoQScatteringDouble} differs in several ways from that given in~\cite{Ozeri2007}. First, we compute $D$-manifold scattering rates directly, whereas~\cite{Ozeri2007} calculates them by multiplying the total scattering rate by the branching ratio to the $D$-manifolds. This overestimates the $D$-scattering rate, as it neglects interference between the $P_{1/2}$ and $P_{3/2}$ manifolds.  Second, we include the V scattering process and the counter-rotating electric field component contribution to the Rabi frequency (\textit{i.e.}, the contribution to Rabi flopping that is neglected in the rotating-wave approximation). Third, we include the detuning dependence of the Lamb-Dicke parameter and scattered photon frequency (while~\cite{Ozeri2007} used a single value of $\eta$ and $\omega_{sc}$ for the entire detuning range). This latter effect makes the scattering rate proportional to $\omega_{sc}^3$, the cube of the scattered photon frequency (Appendix~\ref{appendix:Derivation}). Note that if a detuning is chosen such that $\Delta < -\omega_{Pf}$ the contribution to the scattering probability due to scattering to level $f$ becomes precisely zero. This is because the proportionality to $\omega_{sc}^3$ in such a condition renders the $f$ scattering rate negative, which is non-physical. Another way to see that such scattering is not permitted is to note that it violates energy conservation.

Finally, it is notable that by neglecting the above effects, the simplified model of~\cite{Ozeri2007} implied a lower bound on Raman scattering-induced gate error in $g$ qubits ({$1.06 \times 10^{-4}$}, {$0.50 \times 10^{-4}$}, {$1.46 \times 10^{-4}$}, and {$0.007 \times 10^{-4}$} for Ca${}^+$, Sr${}^+$,  Ba${}^+$, and Yb${}^+$ respectively). This bound appears because the $D$-manifold scattering error in that model approaches a non-zero value at large red detunings. However, in our model, this lower bound does not exist. The Raman scattering error approaches zero at large red detunings. This is as it should be, due to considerations of energy conservation.

\subsection{Higher levels}\label{sec:HigherLevels}

The above models have incorporated scattering probability contributions from higher levels in $m$ qubits, but not $g$ qubits (a disparity we will justify in this section). Higher levels contribute to the scattering probability in two ways: through increasing the overall scattering rate, and increasing or decreasing the Rabi frequency. Because the Rabi frequency and scattering rate have different scaling in detuning, the inclusion of higher levels can actually result in a net decrease in scattering probability for some parameter regimes. 

There are two factors which clearly attenuate the magnitude of contributions from the higher levels to the Raman scattering error: the larger frequency denominators of the scattering rates and the smaller radial overlap with higher levels' wavefunctions. To numerically estimate the contribution of the higher levels, we calculated it for the $P_{3/2}$, $F_{5/2}$, and $F_{7/2}$ manifolds available in the University of Delaware database \cite{UDportal} in Ca${}^+$, Sr${}^+$, and Ba${}^+$.

The contributions from higher levels differ substantially in $m$ qubits and $g$ qubits. Destructive interference in the Rabi frequency due to higher levels causes $m$ qubit behavior to change noticeably at large red- and blue-detunings. Interference effects are largely absent in $g$ qubits, however, since the contributions of higher $P_{1/2}$ and $P_{3/2}$ essentially cancel. The reason for this cancellation is that Raman scattering flips the electron spin via the spin-orbit coupling $\textbf{L} \boldsymbol{\cdot} \textbf{S}$ in the excited state. However, the dipole matrix element for this process is equal in magnitude but opposite in sign for the two fine-structure manifolds $P_{1/2}$ and $P_{3/2}$~\cite{Cline1994,Ozeri2007}. Since we are considering tunings far from the resonant excitation of the higher levels, these matrix elements largely cancel out their contributions; deviations from the $g$ qubit model neglecting higher levels do not exceed more than a few percent until about 1\,PHz detuning, where the laser frequency reaches a resonance with a higher $P$ manifold.

Including the higher levels in the $m$ qubit model lowers scattering probability at the $10^{-4}$ error level by a small amount for red detunings. The corrections for blue detuning, however, can be much larger at the $10^{-4}$ error level; inclusion of the higher levels \textit{increases} the gate error for blue detuning. These corrections are large enough that when higher levels are included, Sr${}^+$ and Ba${}^+$ can no longer get below the $10^{-4}$ error level for blue detuning.

\subsection{Rayleigh scattering errors}\label{sec:OtherRayleigh}

So far, we have neglected Rayleigh scattering-induced errors. Previous discussions of scattering in the literature (\textit{e.g.}~\cite{Ozeri2007}) have characterized the infidelity contribution of Rayleigh scattering-induced dephasing as being proportional to the difference in elastic scattering rates; however, as is shown in~\cite{Uys2010}, the dephasing rate due to Rayleigh scattering must be computed by including interference between the Rayleigh scattering amplitudes. This can lead to Rayleigh scattering becoming the dominant source of error for certain parameter regimes. However, since we are considering clock qubits in both \textit{m} qubits and \textit{g} qubits, the Rayleigh-scattering-induced decoherence will be negligible.~\cite{TingRei2016}

\begin{figure*}[!htb]
    \centering
    \includegraphics[width=1.0\textwidth]{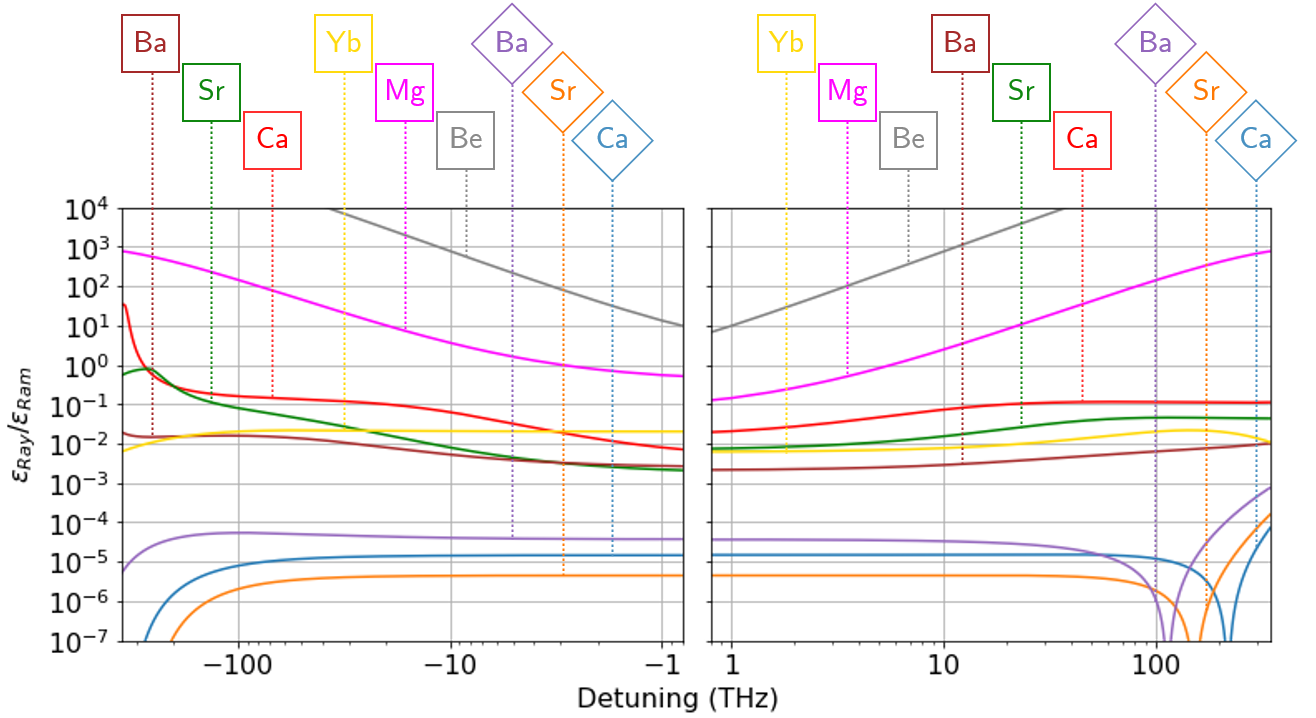}
    \caption{Ratio of the upper bound on Rayleigh recoil error to Raman scattering error during a two-qubit gate for \textit{m} qubits and \textit{g} qubits. The \textit{m} qubit ions are labeled by diamonds and the \textit{g} qubit ions are labeled by squares. Note that, for the detunings plotted, the Rayleigh recoil error upper bound exceeds the Raman scattering error only for \textit{g} qubits in Be${}^+$, Mg${}^+$, and Ca${}^+$. The steep declines in the ratio in \textit{m} qubits at large blue detunings is due to destructive interference with the $F_{7/2}$ manifolds in the Rayleigh scattering rate.}
    \label{fig:eray_eram}
\end{figure*}

Rayleigh scattering can however cause errors during two-qubit gates in two other ways: recoil from the momentum kick during Rayleigh scattering and nonlinearities in the two-qubit gate. Both effects were studied in~\cite{Ozeri2007} and recoil was found to yield a gate error of the form

\begin{equation}\label{eqn:RayleighRecoil}
    \epsilon_{Ray} = P_{E2q} \frac{\langle |\beta|^2 \rangle}{2 K},
\end{equation}
where $P_{E2q}$ is the probability of a elastic Rayleigh scattering event during the two-qubit gate, $\beta$ the recoil displacement in phase space due to the scattering event, and $K$ is the number of loops traced out in phase space (again, {$K = 1$} for our gate). The expected value of $|\beta|^2$ depends on the polarization choice, but we can determine an upper bound on the reduction in fidelity by taking the recoil displacement to equal its maximum value. For our choice of laser beam geometry, the squared magnitude of the recoil displacement is given by~\cite{Ozeri2007}

\begin{equation}\label{eqn:beta_def}
    |\beta|^2 = \frac{\eta^2}{2} \left(\frac{1}{\sqrt{2}} + \text{cos} \theta \right)^2,
\end{equation}
where $\eta$ is the Lamb-Dicke parameter and $\theta$ is the angle of the recoil direction from the axis along the motional mode to which we are coupling. The maximum value is at $\theta = 0$, so

\begin{equation}
    \langle |\beta|^2 \rangle \leq \eta^2 \left(\frac{3}{4} + \frac{1}{\sqrt{2}} \right)
\end{equation}
implying
\begin{equation}\label{eqn:RayleighRecoil}
    \epsilon_{Ray}  \leq P_{E2q} \eta^2 \left(\frac{3}{8} + \frac{1}{2 \sqrt{2}} \right).
\end{equation}
The ratio of this upper bound on Rayleigh recoil error to the Raman scattering error during a two-qubit gate is plotted in Fig.~\ref{fig:eray_eram} for \textit{m} qubits and \textit{g} qubits. The two lightest species, Be${}^+$ and Mg${}^+$, have the largest Rayleigh scattering errors precisely because they are so light; their low masses increase their sensitivity to photon recoil. Additionally, they have higher frequency Raman transitions which makes their Lamb-Dicke parameters larger and therefore further increases their sensitivity to Rayleigh scattering events. In the other species, for most of the detuning range, the Rayleigh recoil errors are small. For \textit{m} qubits, the magnitude of this contribution to the error is 5 or 6 orders of magnitude lower at the $10^{-4}$ error level compared to Raman scattering contribution in the various species of ions we consider. The \textit{g} Rayleigh recoil error is roughly 1 to 3 orders of magnitude smaller than the Raman scattering error for most of the detuning range shown (again, except for Be${}^+$ and Mg${}^+$). The \textit{m} qubits fare better than \textit{g} qubits with respect to the Rayleigh recoil error in large part because \textit{m} qubits have lower branching ratios to the qubit manifold, which leads to a lower probability of elastic scattering events. Furthermore, due to their lower frequency transitions, they have lower Lamb-Dicke parameters, which further suppresses the Rayleigh recoil error.

The infidelity contribution of gate nonlinearities due to recoil momentum displacement is even smaller. As noted in~\cite{Ozeri2007}, the error due to such nonlinearities is proportional to $\eta^4$. This is negligible for most \textit{g} qubits and even less important in \textit{m} qubits, due to their smaller Lamb-Dicke parameters. In Be${}^+$ and Mg${}^+$, this error can still be larger than the Raman scattering error, but it is small compared to the Rayleigh recoil error.

\section{Power Requirements}\label{sec:Power}

Now we will calculate the power required to achieve a given gate error rate (recall that we are considering a two beam single-qubit gate and a three-beam two-qubit gate). We first rewrite Eqn.~\ref{eqn:lifetime},

\begin{equation} \label{eqn:decayrate}
\rho_{q} \frac{\gamma}{g_{Pi}^2} = \frac{4 \hbar \omega^{3}_{3/2}}{3 \pi \epsilon_0 c^3 E^2},
\end{equation} 
where $\rho_q$ is the inelastic fraction of the scattering from $P_{3/2}$ to the qubit manifold, $\omega_{Pi}$ is the frequency of the transition from the qubit manifold to $P_{3/2}$, and $E$ is the peak electric field strength. Note that this equation differs from Eqn. 15 of~\cite{Ozeri2007} only by the factor $\rho_q$. This is because $g_{Pi}$ in~\cite{Ozeri2007} is defined for $S_{1/2} \leftrightarrow P_{3/2}$ transitions, and the branching ratio from $P_{3/2}$ to  the lower $D$ levels was considered small enough that the authors treated the decay rate to $S_{1/2}$ as the total decay rate. In \textit{m} qubits, $g_{Pi}$ is defined for transitions from $D_{5/2} \leftrightarrow P_{3/2}$, but $\rho_q$ is too small to treat this decay rate as the total. The decay rate to an individual level must then be the total decay rate weighted by the branching ratio to that level.

By rewriting $g_{Pi}^2$ in Eqn.~\ref{eqn:decayrate} in terms of $\Omega_R$ and its detuning dependence $r(\Delta)$, defined as

\begin{figure*}[h!]
    \centering
    \includegraphics[width=1.0\textwidth]{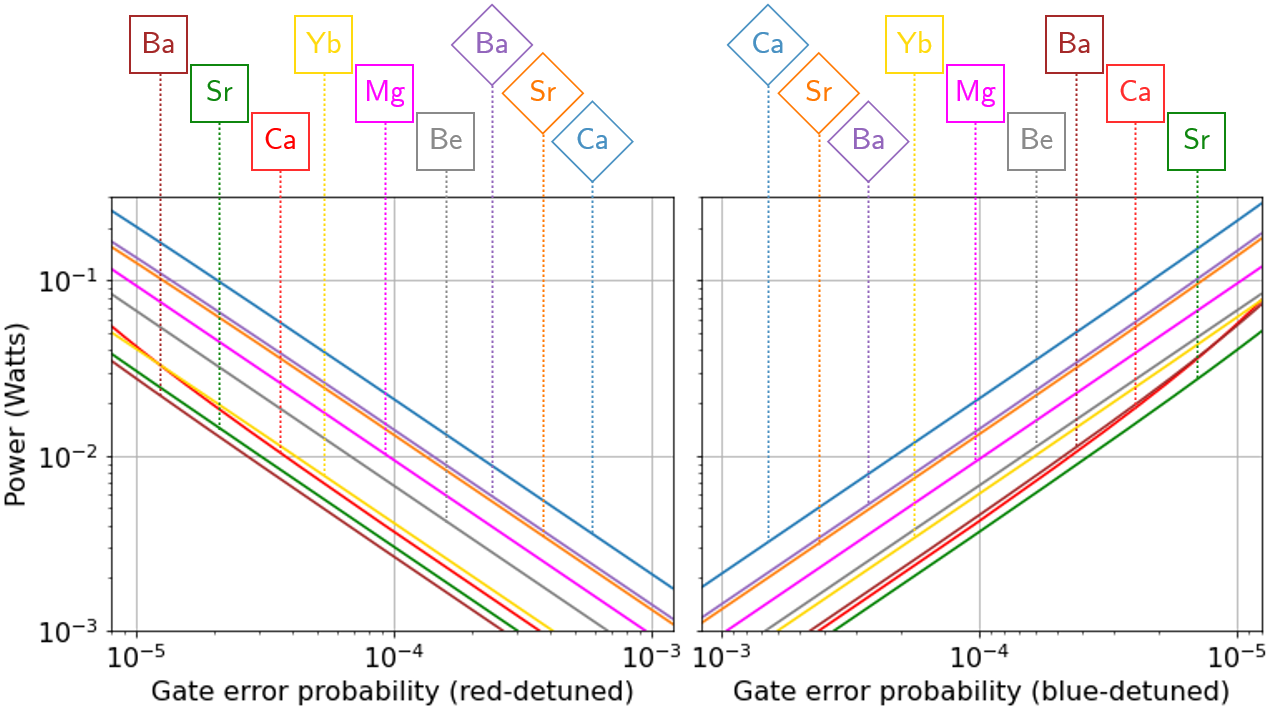}
    \caption{Total power required to achieve a given error rate during a single-qubit $\pi$-rotation gate in the full model (assumes 20 $\mu$m beam waist and 1 $\mu$s gate time). The \textit{m} qubit ions are labeled by diamonds and the \textit{g} qubit ions are labeled by squares. Errors for red-detuned \textit{g} qubits here are calculated from detunings below $P_{1/2}$ only.}
    \label{fig:Pow1q}
\end{figure*}

\begin{figure*}[h!]
    \centering
    \includegraphics[width=1.0\textwidth]{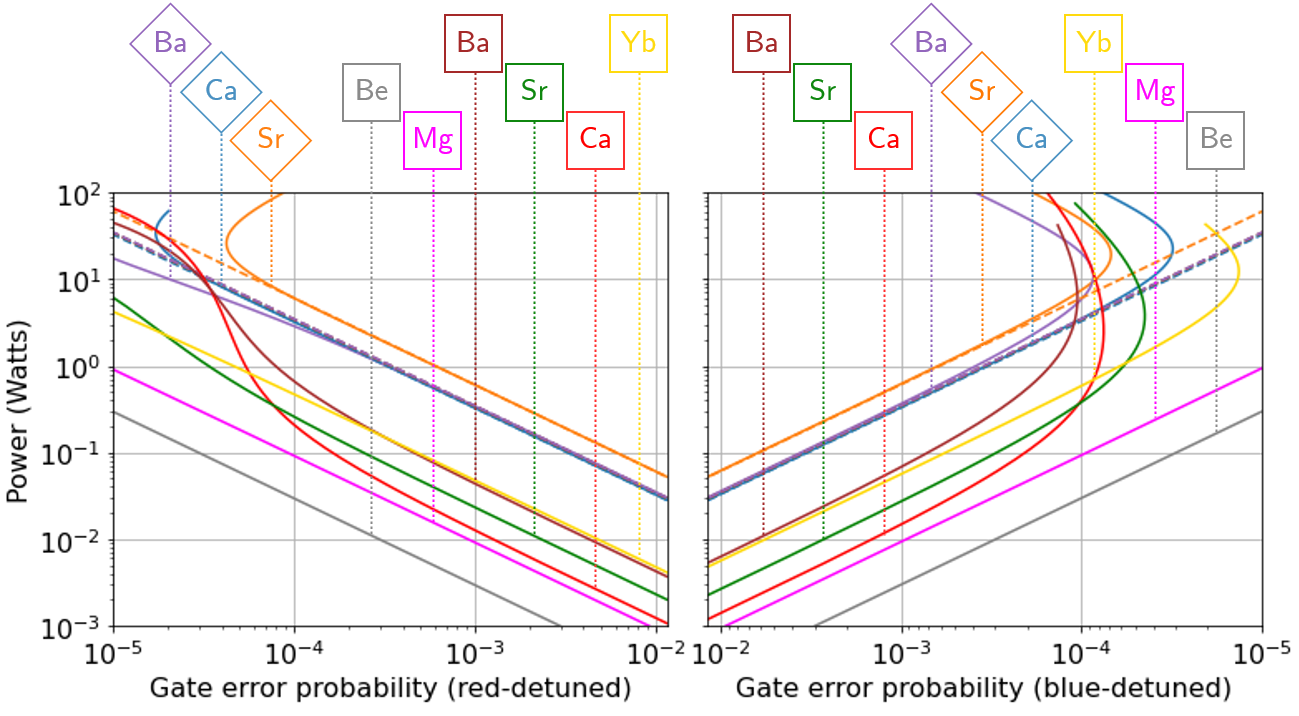}
    \caption{Total power required to achieve a given error rate during a two-qubit M{\o}lmer-S{\o}rensen gate (assumes 20 $\mu$m beam waist, 10 $\mu$s gate time, and a 5\,MHz axial trap frequency). The \textit{m} qubit ions are labeled by diamonds and the \textit{g} qubit ions are labeled by squares. Solid lines correspond to full model, dashed lines correspond to the simplified model. Errors for red-detuned \textit{g} qubits here are calculated from detunings below $P_{1/2}$ only.}
    \label{fig:Pow2qvTot}
\end{figure*}

\begin{widetext}
\begin{equation}
    r(\Delta) = |\Omega_R(\Delta)|/g_{Pi}^2 = \abs*{\sum_k \left(\frac{\langle 1 \rvert \vec{r} \boldsymbol{\cdot} \hat\epsilon_r^* \lvert k \rangle \langle k \rvert \vec{r} \boldsymbol{\cdot} \hat{\epsilon}_b \lvert 0 \rangle}{\mu_{ki}^2 (\omega_{kP} - \Delta)} + \frac{\langle 1 \rvert \vec{r} \boldsymbol{\cdot} \hat\epsilon_r \lvert k \rangle \langle k \rvert \vec{r} \boldsymbol{\cdot} \hat{\epsilon}_b^* \lvert 0 \rangle}{\mu_{ki}^2 (\omega_{ki} + \omega_{Pi} + \Delta)} \right)},
\end{equation}
\end{widetext}
we can write the power requirement as a function of detuning for single and two-qubit gates as (see Appendix~\ref{appendix:Power requirements} for derivation)

\begin{equation} \label{eqn:GeneralPower1q}
\mathscr{P}_{1q}(\Delta) = \frac{\hbar \omega_{Pi}^3 \text{w}_0^2}{3 c^2  \rho_{q} \gamma} \frac{\pi}{\tau_{1q} r(\Delta)}
\end{equation}
and
\begin{equation} \label{eqn:GeneralPower2q}
\mathscr{P}_{2q}(\Delta) = \frac{2 \hbar \omega_{Pi}^3 \text{w}_0^2}{3 \sqrt{2} c^2  \rho_{q} \gamma} \frac{\pi \sqrt{K}}{\tau_{2q} \eta(\Delta) r(\Delta)},
\end{equation}
where $\text{w}_0$ is the beam waist and $\eta(\Delta)$ is the Lamb-Dicke parameter (including detuning dependence). Because the gate error is also some function of $\Delta$, $\epsilon(\Delta)$ (Sections~\ref{sec:SingleQ} and~\ref{sec:TwoQ}), we can visualize the power and error requirements by plotting the set of points $(\epsilon(\Delta), \mathscr{P}(\Delta))$, shown in Figs. \ref{fig:Pow1q} and \ref{fig:Pow2qvTot}.

For the simplified model, $r(\Delta)$ can be directly related to the gate error $\epsilon$, allowing us to write the power as a function of error as (Appendix~\ref{appendix:Power requirements})

\begin{equation} \label{eqn:Perrpi}
\mathscr{P}_{1q}(\epsilon_{1q}) = \rho \frac{5 \pi \hbar \omega_{Pi}^3 \text{w}_0^2}{2 c^2 \epsilon_{1q} } \frac{\pi}{\tau_{1q}}  
\end{equation}
and
\begin{equation} \label{Pow2qgatepi}
\mathscr{P}_{2q}(\epsilon_{2q}) = \rho \frac{10 \pi \hbar \omega_{Pi}^3 \text{w}_0^2}{c^2 \epsilon_{2q}} \frac{\pi}{\tau_{2q}} \frac{ K}{\eta^2}, 
\end{equation}
for single and two-qubit gates, respectively. 

As can be seen in Figs.~\ref{fig:Pow1q} and~\ref{fig:Pow2qvTot}, the power requirements for \textit{g} qubits are, for a large range of detunings, about one order of magnitude smaller than the required power for \textit{m} qubits at the same error probability; this is mostly because of the small branching ratio to $D_{5/2}$ qubit manifold. However, we note that high power lasers are readily available at the large red detunings anticipated for \textit{m} qubits, so these higher power requirements may not be a limiting factor.

From Fig.~\ref{fig:Pow2qvTot}, we can also see the the discrepancies between the simplified model and the full model reappearing. The full model power curves bend backwards past the minima of $\epsilon(\Delta)$, entering a regime of diminishing returns where the gates both take longer and have higher errors.  For a fixed gate time, this farther tuning from resonance means more power is required to drive the gate.

\section{Conclusion}\label{sec:Conclusion}

We constructed a model of two-photon scattering errors during stimulated Raman transitions in trapped-ion qubits which incorporates all two-photon scattering processes, as well as the detuning dependencies of all system parameters involved. We also estimated the contribution to the scattering error from higher energy levels in $m$ qubits and computed an upper bound on Rayleigh scattering error in both $m$ and $g$ qubits, finding this latter error to be negligible in all but the lightest ion species for most of the detuning range considered. We found that including all the above effects produced non-negligible corrections to simpler models of the systems, in particular that the more complete model implies there is no lower bound on Raman scattering-induced infidelity of $g$ qubits as suggested by past models of such errors~\cite{Ozeri2007}.

Additionally, the results of the calculations (summarized in Table~\ref{table:gvsm_comparison}) show that, although $m$ qubits have a lower bound on gate infidelity and the detunings and powers required for $10^{-4}$ error in \textit{m} qubits are larger than in \textit{g} qubits, low errors should still be experimentally achievable in $m$ qubits because high power lasers are more readily available at the required wavelengths for the ion species considered. In sum, low Raman scattering errors are achievable in stimulated Raman-driven gates for both $m$ and $g$ trapped-ion qubits at sufficiently large red detunings.

\section*{Acknowledgements}
We thank Z. Wall for help in checking the accuracy of the calculations. Additionally, we thank J. Chiaverini, J. Gaebler, and M. Foss-Feig for useful discussions. This work was supported in part by the US Army Research Office under award W911NF-20-1-0037. I.D.M., D.J.W., and D.T.C.A. wish to acknowledge support from NSF through the Q-SEnSE Quantum Leap Challenge Institute, Award \#2016244. W.C.C. and E.R.H. acknowledge support from the CIQC
Quantum Leap Challenge Institute through NSF award OMA-2016245 and W.C.C. under award PHY-1912555.

\begin{center}
\begin{table*}[h!]
\setcellgapes{1pt}\makegapedcells
\begin{tabular}{| c | c | c | c | c | c | c | c | c | c | c | } 
\hline
\multicolumn{2}{|c|}{} & ${}^{9}$Be${}^+$ & ${}^{25}$Mg${}^+$ & ${}^{43}$Ca${}^+$ & ${}^{87}$Sr${}^+$ & ${}^{133}$Ba${}^+$ & ${}^{135}$Ba${}^+$ & ${}^{137}$Ba${}^+$ & ${}^{171}$Yb${}^+$ & ${}^{173}$Yb${}^+$\\ 
\hline
\multirow{2}*{\makecell{$\omega_0/2 \pi$ (GHz)}} & \textit{m}  & - & - & 0.025 & 0.036 &  0.062 & 0.012 & 0.00047  & - & -\\\cline{2-11} & \textit{g} & 1.3 & 1.8 & 3.2 & 5.0 &  9.9 & 7.2 & 8.0 & 12.6 & 10.5 \\\cline{2-11}
\hline
\multirow{2}*{\makecell{$\eta$}} & \textit{m}  & - & - & 0.036 & 0.021 & 0.028 & 0.028 & 0.028  & - & -\\\cline{2-11} & \textit{g} & 0.213 & 0.143 & 0.077 & 0.053 & 0.038 & 0.038 & 0.038 & 0.046 & 0.047\\\cline{2-11}

\hline
\multirow{2}*{\makecell{$\Delta/2 \pi$ (THz)}} &  \textit{m} & - & -  & \makecell{ -40.0 \\ (963\,nm)} & \makecell{-66.0 \\ (1335\,nm)} & \makecell{-45.3 \\ (676\,nm)} & \makecell{-45.6 \\ (677\,nm)} & \makecell{-45.9 \\ (677\,nm)}  & - & - \\\cline{2-11} & \textit{g}  & \makecell{ -1.00 \\ (313\,nm)} & \makecell{ -4.55 \\ (281\,nm)}  & \makecell{-9.05 \\ (402\,nm)} & \makecell{-13.0 \\ (429\,nm)} & \makecell{-26.4 \\ (515\,nm)} & \makecell{-26.6 \\ (516\,nm)} & \makecell{-26.9 \\ (516\,nm)} & \makecell{ -15.3 \\ (376\,nm)} & \makecell{ -15.4 \\ (376\,nm)}\\\cline{2-11}
\hline
\multirow{2}*{Power (W)} & \textit{m}  & - & - & 4.9 & 9.1 & 4.4 & 4.4 & 4.5  & - & - \\\cline{2-11} & \textit{g} & 0.067 & 0.13 & 0.30 & 0.37 & 0.94 & 0.96 & 0.98 & 0.67 & 0.67 \\\cline{2-11}
\hline
\end{tabular}
\caption{Comparison of \textit{g} and \textit{m} qubit gate characteristics. The qubit frequency for $m$ and $g$ qubits is given in the first two rows. The Lamb-Dicke parameter $\eta$ is given for a 5\,MHz trap frequency and counter-propagating beams at the  $P_{3/2}$ resonances frequencies (resonance with $D_{5/2}$ and $S_{1/2}$ in \textit{m} and \textit{g} qubits, respectively). The detuning $\Delta$ corresponds to the detuning (from $P_{3/2}$ in $m$ qubits and from $P_{1/2}$ in $g$ qubits) required for $10^{-4}$ error, and the corresponding laser wavelength is given parenthetically below each detuning. Total power requirements are given for the $10^{-4}$ error threshold of the two-qubit M{\o}lmer-S{\o}rensen gate of Section~\ref{sec:TwoQ}, driven by 3 Raman beams.} \label{table:gvsm_comparison}
\end{table*}
\end{center}

\clearpage

\bibliography{refs}
\bibliographystyle{unsrt}

\clearpage

\appendix
\begin{appendices}
\section{Derivation of scattering rate}\label{appendix:Derivation}

\begin{figure}[!htb]
    \centering
    \includegraphics[width=0.4\textwidth]{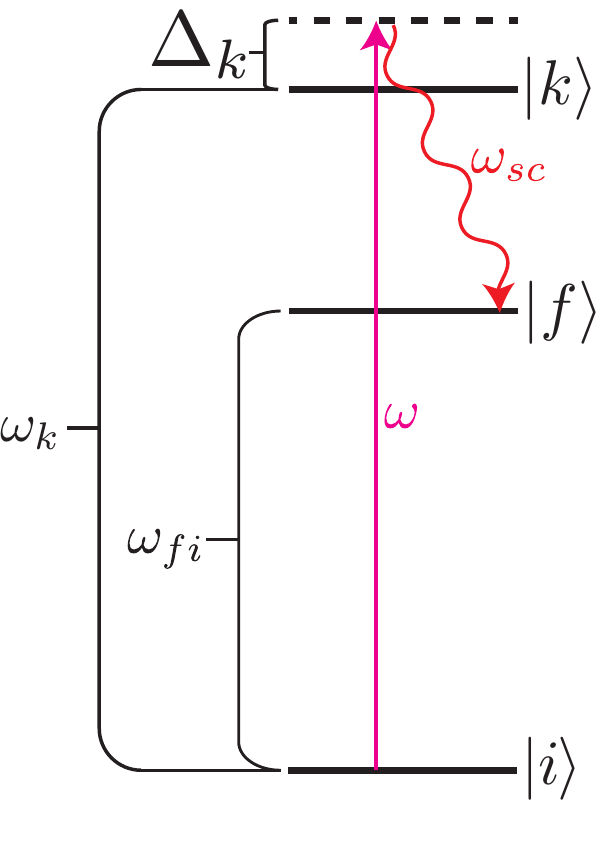}
    \caption{Relevant frequency definitions for Eqn.~\ref{eqn:LoudonRamanRate1}. The transitions are driven by a laser beam of frequency $\omega$. Scattered photon frequency is given by $\omega_{sc} = \omega - \omega_{fi}$.}
    \label{fig:LoudonFreqs}
\end{figure}

To derive the general Raman $\Lambda\!\text{V}$ scattering formula (Eqn.~\ref{eqn:TotalScatteringTwoProcs2_Spon}), begin with Eqn. 8.7.3 of~\cite{Loudon2000},

\hspace{1 cm}

\begin{widetext}
\begin{equation} \label{eqn:LoudonRamanRate1}
\Gamma_{i \rightarrow f, \Lambda\!\text{V}} = \sum_{\boldsymbol{k}_{sc}, \lambda} \frac{\pi e^4 \omega \omega_{sc} n}{2 \epsilon_0^2 \hbar^2 V^2} \Bigl|\sum_k \Bigl( \frac{\bra{f} \vec{r} \boldsymbol{\cdot} \hat{\epsilon}_{\lambda}^* \ket{k} \bra{k} \vec{r} \boldsymbol{\cdot} \hat{\epsilon} \ket{i}}{\omega_k - \omega} + \frac{\bra{f} \vec{r} \boldsymbol{\cdot} \hat{\epsilon} \ket{k} \bra{k} \vec{r} \boldsymbol{\cdot} \hat{\epsilon}_{\lambda}^* \ket{i}}{\omega_k + \omega_{sc}} \Bigr)\Bigr|^2 \delta(\omega_{fi} + \omega_{sc} - \omega).
\end{equation}
\end{widetext}
This equation describes the $\Lambda\!\text{V}$ scattering rate to state $f$ during virtual transitions from state $i$ through the manifold containing the states indexed by $k$ (where all these states are hyperfine sublevels). The transitions are driven by an n-photon laser beam of frequency $\omega$ and polarization $\hat{\epsilon}$, scattering a photon with frequency $\omega_{sc}$ and polarization $\hat{\epsilon}_{\lambda}^*$, where $\lambda$ indexes the two independent polarizations in the chosen basis. The rate is calculated by summing contributions from all scattering modes $\boldsymbol{k}_{sc}, \lambda$ allowed in the quantization volume $V$. The frequencies can be understood by the energy level diagram of Fig.~\ref{fig:LoudonFreqs}.

Rearranging this equation, we find

\begin{widetext}
\begin{equation} \label{eqn:LoudonRamanRate4}
\Gamma_{i \rightarrow f, \Lambda\!\text{V}} = \sum_{\boldsymbol{k}_{sc},\lambda} \frac{\hbar \omega n}{\epsilon_0 V} \frac{\pi e^4 \omega_{sc}}{2 \epsilon_0 \hbar^3 V} \abs*{\sum_k \left( \frac{\bra{f} \vec{r} \boldsymbol{\cdot} \hat{\epsilon}_{\lambda}^* \ket{k} \bra{k} \vec{r} \boldsymbol{\cdot} \hat{\epsilon} \ket{i}}{\omega_k - \omega} + \frac{\bra{f} \vec{r} \boldsymbol{\cdot} \hat{\epsilon} \ket{k} \bra{k} \vec{r} \boldsymbol{\cdot} \hat{\epsilon}_{\lambda}^* \ket{i}}{\omega_k + \omega_{sc}} \right)}^2 \delta(\omega_{fi} + \omega_{sc} - \omega).
\end{equation}
\end{widetext}

Now, if we model the laser field as a classical electric field plane wave with amplitude $E$,
\begin{equation}
    \mathbf{E}_\mathrm{las }(\mathbf{r},t) = \frac{E}{2} \left(\hat{\epsilon} e^{i(\mathbf{k}\cdot \mathbf{r} - \omega t)} + \hat{\epsilon}^\ast e^{-i(\mathbf{k}\cdot \mathbf{r} - \omega t)}  \right),
\end{equation}
we can calculate the temporal and spatial average of
\begin{align}
    \langle \mathbf{E}_\mathrm{las}^\ast(\mathbf{r},t)\boldsymbol{\cdot}\mathbf{E}_\mathrm{las}(\mathbf{r},t) \rangle = \frac{E^2}{2}.
\end{align}
Comparing this to the quantized result in an $n$ photon Fock state, $\bra{n} \hat{E}^\dagger \hat{E} \ket{n} = \frac{\hbar \omega}{\epsilon_0 V}(n+\frac{1}{2})$ allows us to make the replacement, assuming $n \gg 1$,
\begin{equation}
    \frac{2 \hbar \omega}{\epsilon_0 V}n \approx E^2
\end{equation}
to get

\begin{widetext}
\begin{equation}  \label{eqn:LoudonRamanRate5}
\Gamma_{i \rightarrow f, \Lambda\!\text{V}} = \sum_{\boldsymbol{k}_{sc}, \lambda} \frac{E^2 \pi e^4 \omega_{sc}}{4 \epsilon_0 \hbar^3 V} \abs*{\sum_k \left( \frac{\bra{f} \vec{r} \boldsymbol{\cdot} \hat{\epsilon}_{\lambda}^* \ket{k} \bra{k} \vec{r} \boldsymbol{\cdot} \hat{\epsilon} \ket{i}}{\omega_k - \omega} + \frac{\bra{f} \vec{r} \boldsymbol{\cdot} \hat{\epsilon} \ket{k} \bra{k} \vec{r} \boldsymbol{\cdot} \hat{\epsilon}_{\lambda}^* \ket{i}}{\omega_k + \omega_{sc}} \right)}^2 \delta(\omega_{fi} + \omega_{sc} - \omega).
\end{equation}
\end{widetext}





Next, we can take the limit of large quantization volume $V$,
\begin{equation}
    \sum_{\boldsymbol{k}_{sc}} \rightarrow \frac{V}{(2 \pi)^3} \iint d\omega_{sc} d\Omega \frac{\omega_{sc}^2}{c^3},
\end{equation}
enabling us to write the spontaneous scattering rate in the form

\begin{widetext}
\begin{equation}\label{eqn:LoudonRamanRate3}
\begin{split}
    \Gamma_{i \rightarrow f, \Lambda\!\text{V}} = \sum_{\lambda} \iint d\omega_{sc} d\Omega \frac{E^2 e^4 \omega_{sc}^3}{32 \pi^2 c^3 \epsilon_0 \hbar^3} \bigg|\sum_k & \biggl(\frac{\bra{f} \vec{r} \boldsymbol{\cdot} \hat{\epsilon}_{\lambda}^* \ket{k} \bra{k} \vec{r} \boldsymbol{\cdot} \hat{\epsilon} \ket{i}}{\omega_k - \omega} \\ & + \frac{\bra{f} \vec{r} \boldsymbol{\cdot} \hat{\epsilon} \ket{k} \bra{k} \vec{r} \boldsymbol{\cdot} \hat{\epsilon}_{\lambda}^* \ket{i}}{\omega_k + \omega_{sc}} \biggr) \bigg|^2 \delta(\omega_{fi} + \omega_{sc} - \omega).
    \end{split}
\end{equation}
\end{widetext}

Now we wish to perform the integral over ${d\Omega \equiv d\phi \,d\theta \sin(\theta)}$, the direction of the scattered photon's $k$-vector, $\hat{k}_{sc}$.  For this, it is conceptually helpful to gather terms into an expression for the vector transition dipole matrix element for spontaneous scattering,

\begin{widetext}
\begin{align}
   \vec{r}_{sc}(\omega_{sc}) \equiv \frac{E e}{2\hbar} \sum_k \left(\frac{\bra{f} \vec{r}\ket{k} \bra{k} \vec{r} \boldsymbol{\cdot} \hat{\epsilon} \ket{i}}{\omega_k - \omega} + \frac{\bra{f} \vec{r} \boldsymbol{\cdot} \hat{\epsilon} \ket{k} \bra{k} \vec{r} \ket{i}}{\omega_k + \omega_{sc}}\right)
\end{align}
which allows us to write Eq.~(\ref{eqn:LoudonRamanRate3}) in the form
\begin{align}
    \Gamma_{i \rightarrow f, \Lambda\!\text{V}} = \sum_{\lambda} \iint d\omega_{sc} d\Omega \frac{ e^2 \omega_{sc}^3}{8 \pi^2 c^3 \epsilon_0 \hbar} \abs*{  \vec{r}_{sc} \boldsymbol{\cdot} \hat{\epsilon}_{\lambda}^*  }^2 \delta(\omega_{fi} + \omega_{sc} - \omega).
\end{align}
\end{widetext}

Without loss of generality, we will choose a polarization basis such that one of the basis polarization vectors lies in the plane of $\vec{r}_{sc}$ and $\hat{k}_{sc}$; we will call this vector $\hat{\epsilon}_{sc}$ (note that the other basis vector will not contribute to the scattering, as it will necessarily be perpendicular to $\vec{r}_{sc}$). Choosing $\theta$ to be the angle between $\vec{r}_{sc}$ and $\vec{k}_{sc}$, the angle between $\vec{r}_{sc}$ and $\hat{\epsilon}_{sc}$ is $\frac{\pi}{2} - \theta$, which implies

\begin{equation}
\begin{split}
    \abs*{  \vec{r}_{sc} \boldsymbol{\cdot} \hat{\epsilon}_{sc}  }^2 = & |\vec{r}_{sc}|^2 \cos^2\!\left(\mbox{$\frac{\pi}{2}$} - \theta \right) \\ = & |\vec{r}_{sc}|^2\, \sin^2(\theta), 
\end{split}
\end{equation}
allowing us to carry out the integral over the direction of the spontaneously emitted photon to get

\begin{widetext}
\begin{equation}
\begin{split}
    \Gamma_{i \rightarrow f, \Lambda\!\text{V}} =&\,\, \int d\omega_{sc} \frac{ e^2 \omega_{sc}^3}{3 \pi c^3 \epsilon_0 \hbar} \abs*{  \vec{r}_{sc}(\omega_{sc}) }^2 \delta(\omega_{fi} + \omega_{sc} - \omega) \\
    = & \,\,  \frac{ e^2 (\omega - \omega_{fi})^3}{3 \pi c^3 \epsilon_0 \hbar} \abs*{  \vec{r}_{sc}(\omega - \omega_{fi}) }^2 \\
    = & \,\, \frac{ e^2 (\omega - \omega_{fi})^3}{3 \pi c^3 \epsilon_0 \hbar} \left( \left| \vec{r}_{sc}(\omega - \omega_{fi}) \right|\right)^2 \\
    = & \,\,\frac{ e^2 (\omega - \omega_{fi})^3}{3 \pi c^3 \epsilon_0 \hbar} \left( \sqrt{ \sum_q \left|\vec{r}_{sc}(\omega - \omega_{fi}) \boldsymbol{\cdot} \mathbf{\hat{e}}_q\right|^2 }\right)^2\\
    = & \,\,\frac{ e^2 (\omega - \omega_{fi})^3}{3 \pi c^3 \epsilon_0 \hbar} \sum_q \left|\vec{r}_{sc}(\omega - \omega_{fi}) \boldsymbol{\cdot} \mathbf{\hat{e}}_q\right|^2 \label{eqn:post_q_sum}\\
    = & \,\, \frac{ E^2 e^4 (\omega - \omega_{fi})^3}{12 \pi c^3 \epsilon_0 \hbar^3} \sum_q \left| \sum_k \left(\frac{\bra{f} \vec{r} \boldsymbol{\cdot} \mathbf{\hat{e}}_q \ket{k} \bra{k} \vec{r} \boldsymbol{\cdot} \hat{\epsilon} \ket{i}}{\omega_k - \omega} + \frac{\bra{f} \vec{r} \boldsymbol{\cdot} \hat{\epsilon} \ket{k} \bra{k} \vec{r} \boldsymbol{\cdot} \mathbf{\hat{e}}_q \ket{i}}{\omega_k + \omega - \omega_{fi}}\right)\right|^2
\end{split}
\end{equation}
\end{widetext}
where $\mathbf{\hat{e}}_q$ is a polarization vector corresponding to $\pi$, $\sigma^+$, or $\sigma^-$, with $q$ indexing these possibilities.

Now, if $\Delta_k \equiv \omega - \omega_k$ is the detuning relative to some level $k$, we see that we can write $\omega - \omega_{fi}$ as ${\omega_k - \omega_{fi} + \Delta_k}$. Since we choose to measure the detuning relative to the $P_{3/2}$ manifold, we define $\Delta = \omega - \omega_{Pi}$, where $\omega_{Pi}$ is the frequency of the transtion between the manifold containing $\ket{i}$ and $P_{3/2}$. This allows us to rewrite $\omega - \omega_{fi}$ as $\omega_{Pf} + \Delta$, where $\omega_{Pf}$ is the frequency of the transition between the manifold containing state $f$ and $P_{3/2}$; additionally, we may rewrite $\omega_k - \omega$ as $\omega_k - \omega_{Pi} - \Delta$. Rewriting $\omega - \omega_{fi}$ as such and rearranging gives us

\begin{widetext}
\begin{equation} \label{eqn:LoudonRamanRate8}
\begin{split}
\Gamma_{i \rightarrow f, \Lambda\!\text{V}} & = \frac{E^2 e^4 \omega_{Pf}^3}{12 \pi c^3 \epsilon_0 \hbar^3} \sum_q \abs*{\sum_k \left(\frac{\bra{f} \vec{r} \boldsymbol{\cdot} \mathbf{\hat{e}}_q \ket{k} \bra{k} \vec{r} \boldsymbol{\cdot} \hat{\epsilon} \ket{i}}{\omega_k - \omega_{Pi} - \Delta} + \frac{\bra{f} \vec{r} \boldsymbol{\cdot} \hat{\epsilon} \ket{k} \bra{k} \vec{r} \boldsymbol{\cdot} \mathbf{\hat{e}}_q \ket{i}}{\omega_k + \omega_{Pf} + \Delta} \right)}^2 \left(1 + \frac{\Delta}{\omega_{Pf}} \right)^3 \\
& = \frac{e^4 \omega_{Pf}^3}{3 \pi c^3 \epsilon_0 \hbar} \frac{E^2}{4 \hbar^2} \sum_q \abs*{\sum_k \left(\frac{\bra{f} \vec{r} \boldsymbol{\cdot} \mathbf{\hat{e}}_q \ket{k} \bra{k} \vec{r} \boldsymbol{\cdot} \hat{\epsilon} \ket{i}}{\omega_k - \omega_{Pi} - \Delta} + \frac{\bra{f} \vec{r} \boldsymbol{\cdot} \hat{\epsilon} \ket{k} \bra{k} \vec{r} \boldsymbol{\cdot} \mathbf{\hat{e}}_q \ket{i}}{\omega_k + \omega_{Pf} + \Delta} \right)}^2 \left(1 + \frac{\Delta}{\omega_{Pf}} \right)^3.
\end{split}
\end{equation}
\end{widetext}

We begin the next step by noting Eqn.~\ref{eqn:lifetime},

$$\frac{e^2 \omega_{Pf}^3}{3 \pi \epsilon_0 \hbar c^3} \mu_{Pf}^2 = \frac{1}{\tau_{fl}} \equiv \gamma_{Pf}$$
where $\gamma_{Pf}$ is the decay rate from the $P_{3/2}$ manifold to the manifold containing state $f$; this is given by $\alpha_f \gamma$, where $\alpha_f$ is the branching ratio into the manifold containing state $f$ and $\gamma$ is the total spontaneous decay rate for the manifold containing state $l$. With this, we can write

\begin{widetext}
\begin{equation} \label{eqn:LoudonRamanRate10}
\begin{split}
\Gamma_{i \rightarrow f, \Lambda\!\text{V}} & = \frac{\gamma_{Pf}}{\mu_{Pf}^2} \frac{e^2 E^2}{4 \hbar^2} \sum_q \abs*{\sum_k \left(\frac{\bra{f} \vec{r} \boldsymbol{\cdot} \mathbf{\hat{e}}_q \ket{k} \bra{k} \vec{r} \boldsymbol{\cdot} \hat{\epsilon} \ket{i}}{\omega_k - \omega_{Pi} - \Delta} + \frac{\bra{f} \vec{r} \boldsymbol{\cdot} \hat{\epsilon} \ket{k} \bra{k} \vec{r} \boldsymbol{\cdot} \mathbf{\hat{e}}_q \ket{i}}{\omega_k + \omega_{Pf} + \Delta} \right)}^2 \left(1 + \frac{\Delta}{\omega_{Pf}} \right)^3\\
& = \gamma_{Pf} \frac{e^2 E^2}{4 \hbar^2} \sum_q \abs*{\sum_k \left(\frac{\bra{f} \vec{r} \boldsymbol{\cdot} \mathbf{\hat{e}}_q \ket{k} \bra{k} \vec{r} \boldsymbol{\cdot} \hat{\epsilon} \ket{i}}{\mu_{Pf}(\omega_k - \omega_{Pi} - \Delta)} + \frac{\bra{f} \vec{r} \boldsymbol{\cdot} \hat{\epsilon} \ket{k} \bra{k} \vec{r} \boldsymbol{\cdot} \mathbf{\hat{e}}_q \ket{i}}{\mu_{Pf}(\omega_k + \omega_{Pf} + \Delta)} \right)}^2 \left(1 + \frac{\Delta}{\omega_{Pf}} \right)^3.
\end{split}
\end{equation}
\end{widetext}

Now we divide and multiply by $\mu_{Pi}$, \textit{i.e.}, the matrix element of Eqn.~\ref{eqn:mudef} between the $P_{3/2}$ and the manifold containing $i$.

\begin{widetext}
\begin{equation}
\begin{split}
\Gamma_{i \rightarrow f, \Lambda\!\text{V}} =  \gamma_{Pf} \frac{e^2 E^2}{4 \hbar^2} \sum_q \bigg| \sum_k & \mu_{Pi} \biggl(\frac{\bra{f} \vec{r} \boldsymbol{\cdot} \mathbf{\hat{e}}_q \ket{k} \bra{k} \vec{r} \boldsymbol{\cdot} \hat{\epsilon} \ket{i}}{\mu_{Pf} \mu_{Pi} (\omega_k - \omega_{Pi} - \Delta)}\\ & + \frac{\bra{f} \vec{r} \boldsymbol{\cdot} \hat{\epsilon} \ket{k} \bra{k} \vec{r} \boldsymbol{\cdot} \mathbf{\hat{e}}_q \ket{i}}{\mu_{Pf} \mu_{Pi}(\omega_k + \omega_{Pf} + \Delta)} \biggr)\bigg|^2 \left(1 + \frac{\Delta}{\omega_{Pf}} \right)^3 \notag
\end{split}
\end{equation}

\begin{equation} \label{eqn:LoudonRamanRate12}
\begin{split}
\textcolor{white}{\Gamma_{i \rightarrow f, \Lambda\!\text{V}}} = \gamma_{Pf} \frac{e^2 E^2 \mu_{Pi}^2}{4 \hbar^2} \sum_q \bigg| \sum_k & \biggl(\frac{\bra{f} \vec{r} \boldsymbol{\cdot} \mathbf{\hat{e}}_q \ket{k} \bra{k} \vec{r} \boldsymbol{\cdot} \hat{\epsilon} \ket{i}}{\mu_{Pf} \mu_{Pi}(\omega_k - \omega_{Pi} - \Delta)} \\ & + \frac{\bra{f} \vec{r} \boldsymbol{\cdot} \hat{\epsilon} \ket{k} \bra{k} \vec{r} \boldsymbol{\cdot} \mathbf{\hat{e}}_q \ket{i}}{\mu_{Pf} \mu_{Pi} (\omega_k + \omega_{Pf} + \Delta)} \biggr)\bigg|^2 \left(1 + \frac{\Delta}{\omega_{Pf}} \right)^3.
\end{split}
\end{equation}
\end{widetext}
Noting the definition of $g_{Pi}$ (Eqn.~\ref{eqn:gdef}), this becomes

\begin{widetext}
\begin{equation} \label{eqn:LoudonRamanRate14}
\Gamma_{i \rightarrow f, \Lambda\!\text{V}} = \gamma_{Pf} g_{Pi}^2 \sum_q \abs*{\sum_k \left(\frac{\bra{f} \vec{r} \boldsymbol{\cdot} \mathbf{\hat{e}}_q \ket{k} \bra{k} \vec{r} \boldsymbol{\cdot} \hat{\epsilon} \ket{i}}{\mu_{Pf} \mu_{Pi}(\omega_k - \omega_{Pi} - \Delta)} + \frac{\bra{f} \vec{r} \boldsymbol{\cdot} \hat{\epsilon} \ket{k} \bra{k} \vec{r} \boldsymbol{\cdot} \mathbf{\hat{e}}_q \ket{i}}{\mu_{Pf} \mu_{Pi} (\omega_k + \omega_{Pf} + \Delta)} \right)}^2 \left(1 + \frac{\Delta}{\omega_{Pf}} \right)^3.
\end{equation}
\end{widetext}

Finally, we make our result applicable to gates; we do so by averaging over the undisturbed state $\ket{i}$ during the course of the gate. If we consider a $\hat{\sigma}_x$ gate, then an ion initially in the state $\ket{0}$ will get mapped to $\ket{1}$. The ion's state $\ket{i}$ then has a time dependence given by $\ket{i(t)} = \text{cos}(2 \pi t/\tau)\ket{0} + \text{sin}(2 \pi t/\tau)\ket{1}$, where $\tau$ is the gate time. Considering this time dependence and averaging Eqn.~\ref{eqn:LoudonRamanRate14} over the gate time gives

\begin{widetext}
\begin{equation} \label{eqn:LoudonRamanRate15}
\Gamma_{f, \Lambda\!\text{V}} = \gamma_{Pf} \frac{g_{Pi}^2}{2} \sum_{i, q} \abs*{\sum_k \left(\frac{\bra{f} \vec{r} \boldsymbol{\cdot} \mathbf{\hat{e}}_q \ket{k} \bra{k} \vec{r} \boldsymbol{\cdot} \hat{\epsilon} \ket{i}}{\mu_{Pf} \mu_{Pi}(\omega_k - \omega_{Pi} - \Delta)} + \frac{\bra{f} \vec{r} \boldsymbol{\cdot} \hat{\epsilon} \ket{k} \bra{k} \vec{r} \boldsymbol{\cdot} \mathbf{\hat{e}}_q \ket{i}}{\mu_{Pf} \mu_{Pi} (\omega_k + \omega_{Pf} + \Delta)} \right)}^2 \left(1 + \frac{\Delta}{\omega_{Pf}} \right)^3,
\end{equation}
\end{widetext}
where $\ket{i}$ now indexes the two qubit states, $\ket{0}$ and $\ket{1}$.

By the same reasoning, we can get the ladder scattering rate (which contributes in \textit{m} qubits but not \textit{g} qubits),
\begin{widetext}
\begin{equation} \label{eqn:LoudonRamanRateStim}
\begin{split}
\Gamma_{f,\mathrm{lad}} = \gamma_{Pf} \frac{g_{Pi}^2}{2} \sum_{i, q} \bigg| \sum_k & \biggl(\frac{\bra{f} \vec{r} \boldsymbol{\cdot} \mathbf{\hat{e}}_q \ket{k}\bra{k} \vec{r} \boldsymbol{\cdot} \hat{\epsilon}^* \ket{i}}{\mu_{Pf} \mu_{Pi} \left(\omega_{kP} + \Delta + 2 \omega_{PD} \right)} \\  & + \frac{\bra{f} \vec{r} \boldsymbol{\cdot} \hat{\epsilon}^* \ket{k}\bra{k} \vec{r} \boldsymbol{\cdot} \mathbf{\hat{e}}_q \ket{i}}{\mu_{Pf} \mu_{Pi} \left(\omega_{kP} - \Delta + \omega_{Df} \right)}\biggr) \bigg|^2 \left(1 - \frac{2 \omega_{PD} + \Delta}{\omega_{Pf}} \right)^3.
\end{split}
\end{equation}
\end{widetext}

To generate the full model scattering rate, we sum over all scattering events $i \rightarrow f$ except for $i \rightarrow i$ scattering events, since we are ignoring Rayleigh scattering.

We can get the simplified model's scattering rate equation by neglecting the V scattering term in Eqn.~\ref{eqn:LoudonRamanRate15} and assuming $(1 + \Delta/\omega_{Pf})^3 \approx 1$

\begin{equation} \label{eqn:LoudonTotalRate}
\Gamma_{f} \approx \gamma_{Pf} \frac{g_{Pi}^2}{2} \sum_{i} \abs*{\sum_{k,q} \frac{\bra{f} \vec{r} \boldsymbol{\cdot} \mathbf{\hat{e}}_q \ket{k} \bra{k} \vec{r} \boldsymbol{\cdot} \hat{\epsilon} \ket{i}}{\mu_{Pf} \mu_{Pi} \left(\omega_{kP} - \Delta \right)}}^2.
\end{equation}
We can again obtain the Raman scattering rate by summing over all scattering events except for $i \rightarrow i$.

\clearpage
\section{Power requirements derivation}\label{appendix:Power requirements}

Here we derive the expressions given in Section~\ref{sec:Power}. For a Gaussian laser beam of power $\mathscr{P}$ we consider~\cite{steck2020}:
\begin{equation} \label{eqn:E2}
E^2 = \frac{4 \mathscr{P}}{\pi \text{w}_{0}^{2} c \epsilon_0},
\end{equation}

\begin{equation} \label{eqn:decayrateP}
\frac{\gamma}{g_{Pi}^2} = \frac{\hbar \omega_{Pi}^3 \text{w}_0^2}{3 c^2 \mathscr{P} \alpha_q},
\end{equation}
where $\text{w}_0$ is the laser beam waist and $\alpha_q$ is the branching ratio for transitions between $P_{3/2}$ and the qubit manifold. Note that in general $|\Omega_R| = g_{Pi}^2 r(\Delta)$ for some $r(\Delta)$, so we can replace $g_{Pi}^2$ with $|\Omega_R|/r(\Delta)$. This means that the power can be written as

\begin{equation} \label{eqn:GeneralPower}
\mathscr{P} = \frac{\hbar \omega_{Pi}^3 \text{w}_0^2}{3 c^2  \alpha_q \gamma} \frac{|\Omega_R|}{r(\Delta)}.
\end{equation}
For a single-qubit gate, $|\Omega_R| = \pi/2 \tau_{1q}$ where $\tau_{1q}$ is the gate time; for a two-qubit gate, ${|\Omega_R| = \pi \sqrt{K}/2 \sqrt{2} \tau_{2q} \eta(\Delta)}$. This allows us to write the total power required for each as a function of $\Delta$:

\begin{equation} \label{eqn:GeneralPower1q_Appendix}
\mathscr{P}_{1q}(\Delta) = 2 \frac{\hbar \omega_{Pi}^3 \text{w}_0^2}{6 c^2  \alpha_q \gamma} \frac{\pi}{\tau_{1q} r(\Delta)}
\end{equation}
and
\begin{equation} \label{eqn:GeneralPower2q_Appendix}
\mathscr{P}_{2q}(\Delta) = 4 \frac{\hbar \omega_{Pi}^3 \text{w}_0^2}{6 \sqrt{2} c^2  \alpha_q \gamma} \frac{\pi \sqrt{K}}{\tau_{2q} \eta(\Delta) r(\Delta)},
\end{equation}
where the factor of 2 in front of Eqn.~\ref{eqn:GeneralPower1q_Appendix} is due to the use of two beams of equal power, and the factor of 4 in front of Eqn.~\ref{eqn:GeneralPower2q_Appendix} is due to the use of two beams of equal power along with one beam with double the power.

If, as we did in Section \ref{sec:TotalScattering}, we neglect the effects of large detuning, we can write power as a function of the gate error directly. We will start by rewriting $r(\Delta) = 2/15|\Delta|$ in terms of the single-qubit gate error $\epsilon_{1q}$,

\begin{equation} \label{eqn:rD_epsilon}
\epsilon_{1q} = \rho \frac{\pi \gamma}{|\Delta|} = \rho \frac{15 \pi \gamma r(\Delta)}{2} \\
\implies r(\Delta) = \frac{2 \epsilon_{1q}}{15 \rho \pi \gamma},
\end{equation}
or for a two-qubit gate,

\begin{equation} \label{eqn:rD_epsilon_2q}
\begin{split}
\epsilon_{2q} & = \rho \frac{\pi \gamma}{|\Delta|} \frac{4 \sqrt{K}}{\sqrt{2} \eta} = \rho \frac{15 \pi \gamma r(\Delta)}{2} \frac{4 \sqrt{K}}{\sqrt{2} \eta} \\
& \implies r(\Delta) = \frac{2 \epsilon_{2q}}{15 \rho \pi \gamma} \frac{\sqrt{2} \eta}{4 \sqrt{K}}.
\end{split}
\end{equation}

Substituting into Eqns.~\ref{eqn:GeneralPower1q_Appendix} and~\ref{eqn:GeneralPower2q_Appendix}, we get

\begin{equation} \label{eqn:PToterrpi}
\mathscr{P}_{1q}(\epsilon_{1q}) = \rho \frac{5 \pi \hbar \omega_{Pi}^3 \text{w}_0^2}{2 c^2 \epsilon_{1q} \alpha_q} \frac{\pi}{\tau_{1q}}  
\end{equation}
and
\begin{equation} \label{Pow2qgatepi_Appendix}
\mathscr{P}_{2q}(\epsilon_{2q}) = \rho \frac{10 \pi \hbar \omega_{Pi}^3 \text{w}_0^2}{c^2 \epsilon_{2q} \alpha_q} \frac{\pi}{\tau_{2q}} \frac{ K}{\eta^2}. 
\end{equation}

\clearpage

\section{Recovering classical limits}\label{appendix: Recovering classical limits}

We can test the completeness of our model by seeing if it is able to recover classical elastic scattering behavior in the limit of large detuning. Considering only elastic scattering excludes ladder scattering processes (since they are inelastic). This means we can write the elastic scattering rate as

\begin{widetext}
\begin{equation} \label{eqn:ClassicalRate}
    \Gamma = E^2 \frac{e^4 \omega_L^3}{12 \pi c^3 \epsilon_0 \hbar^3} \left| \sum_k |\bra{i} \vec{r} \ket{k}|^2 \left(\frac{1}{\omega_{ki} - \omega_L} +  \frac{1}{\omega_{ki} + \omega_L} \right) \right|^2.
\end{equation}
\end{widetext}

For Raman beams with intensity $I$, we have $E^2 = 2I/\epsilon_0 c$; additionally, we can replace the scattering rate with the scattering cross-section $\sigma$ via $\sigma = \Gamma \hbar \omega_L/I$. Putting these equations together, we get

\begin{widetext}
\begin{equation} \label{eqn:ClassicalCrossSection}
    \sigma = \alpha^2 \frac{8 \pi}{3} \frac{\omega_L^4}{c^2} \left| \sum_k |\bra{i} \vec{r} \ket{k}|^2 \left(\frac{1}{\omega_{ki} - \omega_L} +  \frac{1}{\omega_{ki} + \omega_L} \right) \right|^2,
\end{equation}
\end{widetext}

where $\alpha = e^2/4 \pi \epsilon_0 \hbar c$ is the fine structure constant.

Now we can calculate the limits. For large blue detuning ($\omega_L \gg \omega_{ki}$ for every $k$), we have

\begin{equation} \label{eqn:BlueApprox}
    \left(\frac{1}{\omega_{ki} - \omega_L} +  \frac{1}{\omega_{ki} + \omega_L} \right) \approx -2 \frac{\omega_{ki}}{\omega_L^2},
\end{equation}
allowing us to rewrite Eqn.~\ref{eqn:ClassicalCrossSection} as

\begin{equation} \label{eqn:ClassicalCrossSectionBlue}
    \sigma_\mathrm{blue} = \alpha^2 \frac{32 \pi}{3} \frac{1}{c^2} \left| \sum_k \omega_{ki} |\bra{i} \vec{r} \ket{k}|^2 \right|^2.
\end{equation}
Using the Thomas-Reiche-Kuhn sum rule (for the single valence electron only, and neglecting recoil and the ion's monopole charge), the sum can be evaluated to

\begin{equation} \label{eqn:TRKSumRule}
    \sum_k \omega_{ki} |\bra{i} \vec{r} \ket{k}|^2 = \frac{\hbar}{2 m_e}.
\end{equation}
This gives

\begin{equation} \label{eqn:ClassicalCrossSectionBlueFinal}
    \sigma_\mathrm{blue} = \alpha^2 \frac{32 \pi}{3} \frac{1}{c^2} \frac{\hbar^2}{4 m_e^2} = \frac{8 \pi}{3} \alpha^4 a_0^2,
\end{equation}
where $a_0$ is the Bohr radius and $\alpha^2 a_0$ is the classical electron radius. So for large blue detuning, the model recovers the Thomson cross-section of the valence electron.

For red-detuning, ($\omega_L \ll \omega_{ki}$ for every $k$), we have

\begin{equation} \label{eqn:RedApprox}
    \left(\frac{1}{\omega_{ki} - \omega_L} +  \frac{1}{\omega_{ki} + \omega_L} \right) \approx \frac{2}{\omega_{ki}},
\end{equation}
giving

\begin{equation} \label{eqn:CrossSectionRed}
    \sigma_\mathrm{red} = \alpha^2 \frac{8 \pi}{3} \frac{\omega_L^4}{c^2} \left| \sum_k \frac{2 |\bra{i} \vec{r} \ket{k}|^2}{\omega_{ki}} \right|^2.
\end{equation}
The DC polarizability of the ion can be written as 

\begin{equation} \label{eqn:DCPolarizability}
    \alpha^{(0)} \equiv e^2 \sum_{k \neq i} \frac{\bra{i} \vec{r} \ket{k} \bra{k} \vec{r} \ket{i} + \bra{k} \vec{r} \ket{i} \bra{i} \vec{r} \ket{k}}{E_k - E_i},
\end{equation}
allowing us to rewrite Eqn.~(\ref{eqn:CrossSectionRed}) as

\begin{equation} \label{eqn:ClassicalCrossSectionRed}
\begin{split}
    \sigma_\mathrm{red} & = \frac{8 \pi}{3} \alpha^2 \frac{\omega_L^4}{c^2 e^4} \hbar^2 |\alpha^{(0)}|^2 \\
    & = \frac{8 \pi}{3} \left(\frac{\omega_L}{c}\right)^4 \hbar^2 |\frac{\alpha^{(0)}}{4 \pi \epsilon_0}|^2\\
    & = \frac{8 \pi}{3} k_L^4 \hbar^2 |\alpha^{(0)}|^2.
\end{split}
\end{equation}
We can compare to the classical result by using the Clausius-Mossotti relation,

\begin{widetext}
\begin{equation} \label{eqn:ClausiusMossotti}
    |\alpha^{(0)}|^2 = \left(\frac{3 \epsilon_0}{N} \right)^2 \left( \frac{\epsilon_r - 1}{\epsilon_r + 2} \right) = \left(\frac{3 \epsilon_0}{N} \right)^2 \left( \frac{n^2 - 1}{n^2 + 2} \right) \approx \left(\frac{3 \epsilon_0}{N} \right)^2 \frac{4}{9} (n - 1)^2,
\end{equation}
\end{widetext}
where $N$ is the number density of particles in the material, $\epsilon_r$ is the dielectric constant, and $n$ is the refractive index. The second equality is true for non-magnetic media, and the third, approximate equality holds when $n \approx 1$. Applying this to Eqn.~(\ref{eqn:ClassicalCrossSectionRed}), we find \begin{equation} \label{eqn:ClassicalCrossSectionRedFinal}
    \sigma_\mathrm{red} \approx \frac{2 k_L^4}{3 \pi N^2} |n - 1|^2,
\end{equation}
which is the expression for classical Rayleigh scattering (\textit{e.g.} Jackson~\cite{Jackson1999}). Note that if we had ignored the V scattering process, \textit{i.e.}, the $1/(\omega_{ki} + \omega_L)$ term, we would not have recovered the correct limits.

\end{appendices}

\newpage

\end{document}